\documentclass[conference]{IEEEtran}
\IEEEoverridecommandlockouts

\usepackage{sty/style}

\def\BibTeX{{\rm B\kern-.05em{\sc i\kern-.025em b}\kern-.08em
    T\kern-.1667em\lower.7ex\hbox{E}\kern-.125emX}}
\begin{document}


\title{Joint Beamforming and Matching for \\ Ultra-Dense Massive Antenna Arrays\\
\thanks{$^{*}$equal contribution}
\thanks{For self-consistency, parts of~\fref{sec:modeling_method} intentionally overlap with~\cite[Sec.~II]{stutz_schwan_studer_efficient_and_physically_consistent_modeling_of_reconfigurable_electromagnetic_structures}.}
\thanks{The authors thank Prof.~Hua Wang for providing the scattering parameters of the RF switches used in this work. The work of AST was funded in part by armasuisse. The work of AST, GS, and CS was funded in part by the Swiss State Secretariat for Education, Research, and Innovation (SERI) under the SwissChips initiative, the Swiss National Science Foundation (SNSF) grant 200021\_207314, by a CHIST-ERA grant for the project CHASER (CHIST-ERA-22-WAI-01) through the SNSF grant 20CH21\_218704, and by the European Commission within the context of the project 6G-REFERENCE (6G Hardware Enablers for Cell Free Coherent Communications and Sensing), funded under EU Horizon Europe Grant Agreement 101139155.}
}
\author{\IEEEauthorblockN{Carolina Nolasco-Ferencikova$^\text{$*$,1}$, Georg Schwan$^\text{$*$,1}$, Raphael Rolny$^\text{2}$, Alexander Stutz-Tirri$^\text{1}$, and Christoph Studer$^\text{1}$
\\[0.3cm]
\em $^\text{1}$ETH Zurich, Switzerland; 
$^\text{2}$armasuisse Science and Technology, Switzerland
\\
email: cnolasco@ethz.ch, gschwan@ethz.ch, raphael.rolny@armasuisse.ch, alstutz@ethz.ch, studer@ethz.ch
}
}

\maketitle


\begin{abstract}
Massive multiple-input multiple-output (MIMO) offers substantial spectral-efficiency gains, but scaling to very large antenna arrays with conventional all-digital and hybrid beamforming architectures quickly results in excessively high costs and power consumption. 
Low-cost, switch-based architectures have recently emerged as a potential alternative. However, prior studies rely on simplified models that ignore (among others) antenna coupling, radiation patterns, and matching losses, resulting in inaccurate performance predictions. 
In this paper, we use a physically consistent electromagnetic modeling framework to analyze an ultra-dense patch-antenna array architecture that performs joint beamforming and matching using networks of inexpensive RF switches.
Our results demonstrate that simple, switch-based beamforming architectures can approach the antenna-gain of all-digital solutions at significantly lower cost and complexity. 
\end{abstract}


\section{Introduction}\label{sec:introduction}

Massive multiple-input multiple-output (MIMO) with hundreds of antenna elements can dramatically increase spectral efficiency and reliability in wireless systems~\cite{larsson_massive_MIMO_for_next_generation_wireless_systems}.
However, deploying large antenna arrays remains challenging due to the cost and power consumption of existing beamforming architectures. All-digital solutions that allocate one RF chain per antenna element quickly become expensive and power-hungry as the array size grows; hybrid analog-digital architectures mitigate these issues by reducing the number of RF chains, but still rely on networks of high-resolution phase shifters that remain costly and power hungry~\cite{yan_performance_and_power_tradeoffs_of_beamforming,dutta_a_case_for_digtital_beamforming_at_mmwave}.

In order to address these limitations, recent research has explored hybrid architectures that employ inexpensive RF components, such as RF switches. For example, Eslami Rasekh \emph{et al.}~\cite{rasekh_low_resolution_architectures_for_power_efficient_scaling_of_mmWave_phased_array_receivers} propose replacing conventional phase shifters with low-resolution, switch-based binary phase shifters, offering a viable approach towards low-cost and energy-efficient beamforming.
Despite their promise, existing research on such low-cost architectures relies on simplified and physically inconsistent models that neglect real-world effects, including mutual coupling, non-ideal element patterns, ohmic losses, and matching losses.
These effects become particularly important with inexpensive and highly constrained RF hardware, and ignoring them leads to inaccurate performance predictions.
As a result, the real-world efficacy of such low-cost architectures is currently unclear.

\subsection{Contributions}
In this paper, we demonstrate that physically consistent modeling enables the systematic design, optimization, and analysis of novel beamforming architectures capable of scaling MIMO arrays to extreme dimensions.
Specifically, we utilize the framework introduced in~\cite[Sec.~II]{stutz_schwan_studer_efficient_and_physically_consistent_modeling_of_reconfigurable_electromagnetic_structures} to model an ultra-dense massive patch-antenna array, capturing mutual coupling, antenna element patterns, losses, etc. 
We then propose a new hybrid beamforming architecture that performs joint beamforming and matching using low-cost RF components (i.e., networks of inexpensive RF switches and transmission lines), thereby eliminating the need for costly and power-hungry phase shifters or other analog circuitry.
Our results confirm that simple beamforming architectures can approach the antenna gain of all-digital architectures at significantly lower costs.

\subsection{Notation}
We use lowercase boldface for general vectors (e.g.,~$\vect{a}$) and uppercase boldface for general matrices (e.g., $\mat{A}$). 
We use pink sans-serif (e.g.,~$\phs{a}$) and pink sans-serif boldface (e.g.,~$\phv{a}$) for phasors (cf.~\cite[Def.~1]{stutz_schwan_studer_efficient_and_physically_consistent_modeling_of_reconfigurable_electromagnetic_structures}) and vectors containing phasors, respectively. 
The superscripts $^\T$ and $^\He$ indicate transpose (e.g.,~$\mat{A}^\T$) and conjugate transpose (e.g.,~$\mat{A}^\He$), respectively.
Given a vector $\vect{a}$, we use $\diag(\vect{a})$ to indicate the diagonal matrix with the elements of $\vect{a}$ in the main diagonal.
We denote the Euclidean norm by $\|\cdot\|_2$. 
We use blackboard bold font for operators (e.g.,~$\mathbb{S}$). 
For a complex number $z\in\mathbb{C}$, the real part is $\Re\{z\}$.
Given $N\in\mathbb{N}$, we define the set $[N]\triangleq\{1,\ldots,N\}$.
\def\waves at (#1,#2,#3,#4,#5){
    \draw [shift={(#1,#2)}, line width=0.5pt, color=#5, arrows = {-Stealth[inset=0, length=5pt, angle'=35]}] (-.5,.2) -- (.5,.2);

    \node[shift={(#1,#2)}, anchor=north] at (0,-.2+.02) {\footnotesize #3};

    \draw [shift={(#1,#2)}, line width=0.5pt, color=#5, arrows = {-Stealth[inset=0, length=5pt, angle'=35]}] (.5,-.2) -- (-.5,-.2);

    \node[shift={(#1,#2)}, anchor=north] at (0,.2+.06) {\footnotesize #4};
   
}

\begin{figure}[t!]
    \centering
    {
    \small
    \begin{tikzpicture}
        
        \def\distDown{2};
        
        \def\voltageX{2.};
        
        \def\impedanceX{2.7};
        
        \def\firstBoundX{3.9};
        \def\stCenterX{4.9};
        \def\stWidth{1.2};
        
        \def\secondBoundX{6.3};
        
        \def\srCenterX{7.3};
        \def\srWidth{1.2};
        
        \def\thirdBoundx{8.2};

        \def\waveLengthMiddleX{9.3};
        \def\waveScaleX{1};
        \def\waveScaleY{1};

        \node[shift={(\firstBoundX/2+1/2,0)}, anchor=north, align=center] at (0,5.2) {\normalsize RF frontend};
        
        \draw[shift={(\firstBoundX,0)}, line width=.7, dashed, dash pattern=on 6pt off 3pt] (0,5.2) -- (0,4.1);
        \draw[shift={(\firstBoundX,0)}, line width=.7, dashed, dash pattern=on 6pt off 3pt] (0,2.9) -- (0,4.1-\distDown);
        
        \draw[shift={(\firstBoundX,0)}, line width=.7, dashed, dash pattern=on 6pt off 3pt] (0,2.9-\distDown) -- (0,2.5-\distDown);
        
        \draw[shift={(\secondBoundX,0)}, line width=.7, dashed, dash pattern=on 6pt off 3pt] (0,5.2) -- (0,4.1);
        \draw[shift={(\secondBoundX,0)}, line width=.7, dashed, dash pattern=on 6pt off 3pt] (0,2.9) -- (0,4.1-\distDown);
        
        \draw[shift={(\secondBoundX,0)}, line width=.7, dashed, dash pattern=on 6pt off 3pt] (0,2.9-\distDown) -- (0,2.5-\distDown);
        
        \draw[shift={(\thirdBoundx,0)}, line width=.7, dashed, dash pattern=on 6pt off 3pt] (0,5.2) -- (0,2.5-\distDown);

        \begin{scope}[shift={(-0.1,-0.1)}]
            \node[shift={(\firstBoundX,2.8-.2)}, anchor=west, fill=white] at (-.2,0) {\color{grey} \footnotesize ($N$)};
            \draw  [shift={(\firstBoundX,2.8-.2)}, line width=.5pt, fill=grey, color=grey] (-.2,.14)  ellipse (0.02 and 0.02);
            \draw  [shift={(\firstBoundX,2.8-.2)}, line width=.5pt, fill=grey, color=grey] (-.2,0) ellipse (0.02 and 0.02);
            \draw  [shift={(\firstBoundX,2.8-.2)}, line width=.5pt, fill=grey, color=grey] (-.2,-0.14) ellipse (0.02 and 0.02);
        \end{scope}

        \begin{scope}[shift={(-0.1,-0.1)}]
            \node[shift={(\secondBoundX,2.8-.2)}, anchor=west, fill=white] at (-.2,0) {\color{grey} \footnotesize ($M$)};
            \draw  [shift={(\secondBoundX,2.8-.2)}, line width=.5pt, fill=grey, color=grey] (-.2,.14)  ellipse (0.02 and 0.02);
            \draw  [shift={(\secondBoundX,2.8-.2)}, line width=.5pt, fill=grey, color=grey] (-.2,0) ellipse (0.02 and 0.02);
            \draw  [shift={(\secondBoundX,2.8-.2)}, line width=.5pt, fill=grey, color=grey] (-.2,-0.14) ellipse (0.02 and 0.02);
        \end{scope}

        \waves at (\firstBoundX,2.5-0.5-\distDown,$\phv{b}_\up{T}$,$\phv{a}_\up{T}$,black);
        \waves at (\secondBoundX,2.5-0.5-\distDown,$\phv{b}_\up{R}$,$\phv{a}_\up{R}$,black);

        \draw [shift={((\thirdBoundx,2.5-0.5-\distDown))}, line width=0.5pt, color=black, arrows = {-Stealth[inset=0, length=5pt, angle'=35]}] (-.5,.2) -- (.5,.2);
    
        \node[shift={((\thirdBoundx,2.5-0.5-\distDown))}, anchor=north] at (0,.2+.02) {\footnotesize $\phv{a}_\up{F}$};

        \node[shift={(\firstBoundX/2+\secondBoundX/2,0)}, anchor=north, align=center] at (0,5.2) {\normalsize tuning\\ \normalsize network};

        \node[shift={(\secondBoundX/2+\thirdBoundx/2,0)}, anchor=north, align=center] at (0,5.2) {\normalsize radiating\\ \normalsize structure};

        \node[shift={(\waveLengthMiddleX,0)}, anchor=north] at (-.25,5.2) {\normalsize far field};

        \draw  [shift={(\thirdBoundx,3.6-\distDown/2)}, anchor=center, fill=white, color=white] (-0.3,-0.3) rectangle (.3,.3);

        \draw [shift={(\thirdBoundx,3.6-\distDown/2)}, line width=.7pt] plot[smooth, tension=.8] coordinates {(0,0) (.15\waveScaleX,0) (.3\waveScaleX,.025*2) (.45\waveScaleX,-.1\waveScaleY) (.6\waveScaleX,.175\waveScaleY) (.75\waveScaleX,-.175\waveScaleY) (.9\waveScaleX,.1\waveScaleY) (1.05\waveScaleX,-.025\waveScaleY) (1.2\waveScaleX,0) (1.35\waveScaleX,.0) };
        \draw [shift={(\thirdBoundx,3.6-\distDown/2)}, color=black, arrows = {-Stealth[inset=0, length=5pt, angle'=35]}] (1.35\waveScaleX,0) -- (1.45\waveScaleX,0);

        \node[shift={(\voltageX, 3.5)}, anchor=center] at (-.7,-.025) {\footnotesize $\phs{v}_{\up{Tx},1}$};
        \node[shift={(\impedanceX, 3.9)}, anchor=south] at (0,0.05) {\footnotesize $Z_{\up{Tx},1}$};
        \node[shift={(\firstBoundX, 3.9)}, anchor=south west] at (-.85,0) {\footnotesize $\phs{i}_{\up{T},1}$};
        \node[shift={(\firstBoundX, 3.5)}, anchor=center] at (0,-.025) {\footnotesize $\phs{v}_{\up{T},1}$};
        \node[shift={(\secondBoundX, 3.9)}, anchor=south] at (-0.4,0) {\footnotesize $\phs{i}_{\up{R},1}$};
        \node[shift={(\secondBoundX, 3.5)}, anchor=center] at (0,-.025) {\footnotesize $\phs{v}_{\up{R},1}$};
        \node[shift={(0,2.9)}, anchor=north west] at (0.7,0) { \color{grey} PA $1$};

        \begin{scope}[shift={(0,-\distDown)}]
            \node[shift={(\voltageX, 3.5)}, anchor=center] at (-.7,-.025) {\footnotesize $\phs{v}_{\up{Tx},N}$};
            \node[shift={(\impedanceX, 3.9)}, anchor=south] at (0,0.05) {\footnotesize $Z_{\up{Tx},N}$};
            \node[shift={(\firstBoundX, 3.9)}, anchor=south west] at (-.85,0) {\footnotesize $\phs{i}_{\up{T},N}$};
            \node[shift={(\firstBoundX, 3.5)}, anchor=center] at (0,-.025) {\footnotesize $\phs{v}_{\up{T},N}$};
            \node[shift={(\secondBoundX, 3.9)}, anchor=south] at (-0.4,0) {\footnotesize $\phs{i}_{\up{R},M}$};
            \node[shift={(\secondBoundX, 3.5)}, anchor=center] at (0,-.025) {\footnotesize $\phs{v}_{\up{R},M}$};
            \node[shift={(0,2.9)}, anchor=north west] at (0.7,0) { \color{grey} PA $N$};
        \end{scope}
        
        \foreach \y in {0, -\distDown} {
            \begin{scope}[shift={(0,\y)}]

            \draw[line width=.6] (\voltageX,3.9) -- (\secondBoundX/2+\thirdBoundx/2,3.9);
            \draw[line width=.6] (\voltageX,3.1) -- (\secondBoundX/2+\thirdBoundx/2,3.1);
            
            \draw [shift={(\secondBoundX-0.5, 3.9)}, line width=0.3pt, arrows = {-Stealth[inset=0, length=6pt, angle'=25]}] (-.1,0) -- (.1,0);
    
            \node[shift={(\secondBoundX, 3.5)}, anchor=center] at (0,.2) {\scriptsize+};
            \node[shift={(\secondBoundX, 3.5)}, anchor=center] at (0,-.25) {\scriptsize--};

            \draw  [shift={(\secondBoundX,3.9)}, line width=.6pt, fill=white] (0,0) ellipse (0.09 and 0.09);
            \draw  [shift={(\secondBoundX,3.9-.8)}, line width=.6pt, fill=white] (0,0) ellipse (0.09 and 0.09);

            \draw[line width=.6] (2,3.1) -- (2,3.9);

            \draw  [line width=.7pt, dashed, color=grey] (.8,2.9) rectangle (3.1,4.42);
            
            \draw  [shift={(\firstBoundX,3.9)}, line width=.6pt, fill=white] (0,0) ellipse (0.09 and 0.09);
            \draw  [shift={(\firstBoundX,3.9-.8)}, line width=.6pt, fill=white] (0,0) ellipse (0.09 and 0.09);

    
            \draw  [shift={(\impedanceX,3.9)}, line width=.7pt, anchor=center, fill=white] (-.3,-.12) rectangle (.3,.12);
    
            \draw[shift={(\voltageX,3.5)}, line width=.7pt] (0,0) ellipse (0.28 and 0.28);
            \draw[shift={(\voltageX,3.5)}, line width=.7pt] (0,-.28) -- (0,.28);
            
            \node[shift={(\voltageX, 3.5)}, anchor=center] at (-.5,.2) {\scriptsize+};
            \node[shift={(\voltageX, 3.5)}, anchor=center] at (-.5,-.25) {\scriptsize--};
            
            \draw [shift={(\firstBoundX-0.5, 3.9)}, line width=0.3pt, arrows = {-Stealth[inset=0, length=6pt, angle'=25]}] (-.1,0) -- (.1,0);
    
            \node[shift={(\firstBoundX, 3.5)}, anchor=center] at (0,.2) {\scriptsize+};
            \node[shift={(\firstBoundX, 3.5)}, anchor=center] at (0,-.25) {\scriptsize--};

            
            \end{scope}
        }

        \draw  [shift={(\stCenterX,0)}, line width=.7pt, anchor=center, fill=white] (-\stWidth/2,2.8-\distDown) rectangle (\stWidth/2,4.2);
        \node[shift={(\stCenterX,0)}, anchor=center] at (0,3.5-\distDown/2) {\Large $\mat{S}_\up{T}$};

        \draw  [shift={(\srCenterX,0)}, line width=.7pt, anchor=center, fill=white] (-\srWidth/2,2.8-\distDown) rectangle (\srWidth/2,4.2);
        \node[shift={(\srCenterX,0)}, anchor=center] at (0,3.5-\distDown/2) {\Large $\oper{S}_\up{R}$};

    \end{tikzpicture}}
    \vspace{-.5cm}
\caption{
Structure of the model used in this paper. The \emph{RF frontend} represents $N$ power amplifiers; the \emph{tuning network} represents reconfigurable analog beamforming and matching RF circuitry; and the \emph{radiating structure} represents 
$M$ antenna elements. 
The original structure shown in~\cite[Fig.~2]{stutz_schwan_studer_efficient_and_physically_consistent_modeling_of_reconfigurable_electromagnetic_structures} additionally models low-noise amplifiers (for receivers), noise from various sources, and incoming far-field waves---all of which are not used here.}
\label{fig:basic_structure}
\end{figure}

\section{Modeling Method}\label{sec:modeling_method}

\subsection{REMS Model}
In~\cite[Sec.~II]{stutz_schwan_studer_efficient_and_physically_consistent_modeling_of_reconfigurable_electromagnetic_structures}, we proposed an \emph{efficient} and \emph{physically consistent} far-field modeling framework for \emph{general reconfigurable electromagnetic structures (REMSs)}. Throughout this paper, we will utilize parts of this model for our analysis.

With \emph{efficient}, we mean that (i) the model parameters can be obtained with acceptable computational effort at design time, and (ii) the model enables real-time prediction of a REMS's behavior with very low complexity. 
Concretely, our approach allows each REMS in a wireless communication or sensing system to be characterized separately using a \emph{single} full-wave EM simulation or measurement campaign. 
These characterizations can then be used to quickly compute the system-level behavior for arbitrary REMS positions, RF circuitry configurations, and RF frontend output signals. 

With \emph{physically consistent}, we mean that our model’s predictions adhere to the physical laws dictated by Maxwell’s equations. 
This includes effects such as mutual (inter-antenna) coupling, nonreciprocal materials, polarization, ohmic and matching losses, the influence of metallic housing, noise contributed by low-noise amplifiers, and noise generated within or received by antennas. 
Furthermore, when predicting the properties of wireless channels, our approach is significantly more numerically stable and, consequently, more accurate, while running orders of magnitude faster than full-wave EM simulations, as we demonstrate in~\cite[Sec.~III]{stutz_schwan_studer_efficient_and_physically_consistent_modeling_of_reconfigurable_electromagnetic_structures}.\footnote{A detailed discussion on numeric stability and required computational resources can be found in~\cite[Sec.~III-C]{stutz_schwan_studer_efficient_and_physically_consistent_modeling_of_reconfigurable_electromagnetic_structures}.}

With \emph{general REMSs}, we refer to a wide class of wireless communication or sensing systems, including all-digital\footnote{We speak of an \emph{all-digital} architecture when the number of RF chains $N$ is equal to the number of antenna elements $M$.} or hybrid multi-antenna systems, reconfigurable reflectarrays, and passive or active reconfigurable intelligent surfaces. 
Practically, almost any system with a linear, time-invariant \emph{radiating structure} can be described by our model.\footnote{To be specific, any system for which Assumptions~1--3 in~\cite{stutz_schwan_studer_efficient_and_physically_consistent_modeling_of_reconfigurable_electromagnetic_structures} hold can be described by our model.} 

\subsection{Aspects of the REMS Model Used in This Work}
We now review the components of our REMS model that we use in~\fref{sec:case_study}.\footnote{To maintain consistency, in this paper, we repeat selected material from~\cite{stutz_schwan_studer_efficient_and_physically_consistent_modeling_of_reconfigurable_electromagnetic_structures}.} 

Our model characterizes the REMS’s behavior at an arbitrary frequency $f\in\mathbb{R}_{>0}$.\footnote{This does not limit our model to narrowband analysis; it simply means that if multiple frequencies are of interest, the modeling procedure must be applied separately to each one.}
For a given frequency $f$, we denote the free space wavenumber, impedance, and wavelength by~\mbox{$k\triangleq 2\pi f\sqrt{\mu_\up{0}\varepsilon_\up{0}}$}, $Z_\up{0}\triangleq\sqrt{\mu_\up{0}/\varepsilon_\up{0}}$, and $\lambda\triangleq (\sqrt{\varepsilon_\up{0} \mu_\up{0}}f)^{-1}$ respectively,  where $\mu_\up{0}$ is the permeability and $\varepsilon_\up{0}$ is the permittivity of free space. 
We use the physicist's convention for spherical coordinate systems~\cite{ISO_quantities_and_units_2_mathematics}, with $r$ as the \emph{radial distance}, $\theta$ as the \emph{polar angle}, and $\varphi$ as the \emph{azimuthal angle}.
Furthermore, for each coordinate $(r,\,\theta,\,\varphi)$, we denote the local orthogonal unit vectors in the directions of increasing $r$, $\theta$, and~$\varphi$ as~$\hat{\vect{r}}$,~$\hat{\vect{\theta}}$, and~$\hat{\vect{\varphi}}$, respectively. 
Finally, we denote the set of all angle pairs $(\theta,\varphi)$ on the unit sphere by $\Omega\triangleq[0,\pi]\times[0,2\pi)$. 

To describe the internal behavior of a REMS, we rely on a circuit-theoretic approach based on scattering parameters and \emph{circuit-theoretic power waves} as defined in~\cite[Def.~2]{stutz_schwan_studer_efficient_and_physically_consistent_modeling_of_reconfigurable_electromagnetic_structures}.
%

\def\vertT at (#1,#2,#3) named (#4){

    \begin{scope}[shift={(#1,#2)}]
        \draw[white, fill=white, line width=1.15] (0,0) circle (.2);
        \draw[line width=1.15, pattern=diagonal pattern node ,pattern color=#3] (0,0) circle (.2);
        \node at (0,.5) {\normalsize #4};
    \end{scope}
}
\def\vertB at (#1,#2,#3) named (#4){

    \begin{scope}[shift={(#1,#2)}]
        \draw[white, fill=white, line width=1.15] (0,0) circle (.2);
        \draw[line width=1.15, pattern=diagonal pattern node, pattern color=#3] (0,0) circle (.2);
        \node at (0,-.5) {\normalsize #4};
    \end{scope}
}
\def\vertL at (#1,#2,#3) named (#4){

    \begin{scope}[shift={(#1,#2)}]
        \draw[white, fill=white, line width=1.15] (0,0) circle (.2);
        \draw[line width=1.15, pattern=diagonal pattern node, pattern color=#3] (0,0) circle (.2);
        \node at (-.5,0) {\normalsize #4};
    \end{scope}
}

\def\edgeL from (#1,#2) to (#3,#4) named (#5){
    \draw [line width=1pt, color=blue_120, arrows = {-Stealth[inset=0, length=8pt, angle'=35]}] (#1/2+#3/2-.4,#2/2+#4/2) -- (#1/2+#3/2-.4,#2/2+#4/2-4pt);
    \draw [line width=1pt, color=blue_120] plot[smooth, tension=1.2] coordinates {(#1,#2) (#1/2+#3/2-.4,#2/2+#4/2) (#3,#4)};
    \node[anchor=east] at (#1/2+#3/2-.4,#2/2+#4/2) {\color{blue_120}\normalsize #5};
}

\def\edgeR from (#1,#2) to (#3,#4) named (#5){
    \draw [line width=1pt, color=blue_120, arrows = {-Stealth[inset=0, length=8pt, angle'=35]}] (#1/2+#3/2+.4,#2/2+#4/2) -- (#1/2+#3/2+.4,#2/2+#4/2+4pt);
    \draw [line width=1pt, color=blue_120] plot[smooth, tension=1.2] coordinates {(#1,#2) (#1/2+#3/2+.4,#2/2+#4/2) (#3,#4)};
    \node[anchor=west] at (#1/2+#3/2+.45,#2/2+#4/2) {\color{blue_120}\normalsize #5};
}

\def\edgeT from (#1,#2) to (#3,#4) named (#5){
    \draw [line width=1pt, color=blue_120, arrows = {-Stealth[inset=0, length=8pt, angle'=35]}] (#1/2+#3/2,#2/2+#4/2-.4) -- (#1/2+#3/2-.1,#2/2+#4/2-.4);
    \draw [line width=1pt, color=blue_120] plot[smooth, tension=1.2] coordinates {(#1,#2) (#1/2+#3/2,#2/2+#4/2-.4) (#3,#4)};
    \node[anchor=north] at (#1/2+#3/2,#2/2+#4/2-.4) {\color{blue_120}\normalsize #5};
}

\def\edgeB from (#1,#2) to (#3,#4) named (#5){
    \draw [line width=1pt, color=blue_120, arrows = {-Stealth[inset=0, length=8pt, angle'=35]}] (#1/2+#3/2,#2/2+#4/2+.4) -- (#1/2+#3/2+.1,#2/2+#4/2+.4);
    \draw [line width=1pt, color=blue_120] plot[smooth, tension=1.2] coordinates {(#1,#2) (#1/2+#3/2,#2/2+#4/2+.4) (#3,#4)};
    \node[anchor=south] at (#1/2+#3/2,#2/2+#4/2+.4) {\color{blue_120}\normalsize #5};
}

\def\edgeRS from (#1,#2) to (#3,#4) named (#5){
    \draw [line width=1pt, color=blue_120, arrows = {-Stealth[inset=0, length=8pt, angle'=35]}] (.25,.55*\distV) -- (0.25,.6*\distV);
    \draw [line width=1pt, color=blue_120] plot[smooth, tension=1.2] coordinates {(#1,#2) (#1/2+#3/2+.25,#2/2+#4/2) (#3,#4)};
    \node[anchor=east] at (#1/2+#3/2+.2+.05,#2/2+#4/2-.05) {\color{blue_120}\normalsize #5};
}

\begin{figure}[tp]
    \centering
    {
    \small
    \begin{tikzpicture}
        \def\distH{2.6};
        \def\distV{2.3};

        \clip (-.8,-.8*\distV) rectangle (3.6*\distV,1*\distV);

        \edgeL from (\distH,\distV/2) to (\distH,-\distV/2) named ($\mat{S}_\up{RF}$);
        \edgeL from (2*\distH,\distV/2) to (2*\distH,-\distV/2) named ($\mat{S}_{\up{T}_\up{RR}}$);

        \edgeR from (\distH,\distV/2) to (\distH,-\distV/2) named ($\mat{S}_{\up{T}_\up{TT}}$);
        \edgeR from (2*\distH,\distV/2) to (2*\distH,-\distV/2) named ($\oper{S}_{\up{R}_\up{RR}}$);

        \edgeT from (2*\distH,\distV/2) to (\distH,\distV/2) named ($\mat{S}_{\up{T}_\up{TR}}$);

        \edgeB from (2.7*\distH,-\distV/2) to (2*\distH,-\distV/2) named ($\oper{S}_{\up{R}_\up{FR}}$);
        \edgeB from (2*\distH,-\distV/2) to (1*\distH,-\distV/2) named ($\mat{S}_{\up{T}_\up{RT}}$);
        
        \edgeB from (1*\distH,-\distV/2) to (0.3*\distH,-\distV/2) named ($\mat{K}_{\phv{v}_\up{Tx}}$);

        \vertL at (0.3*\distH,-\distV/2,white) named ($\phv{v}_\up{Tx}$);
    
        \vertT at (\distH,\distV/2,white) named ($\phv{b}_\up{T}$);
        \vertB at (\distH,-\distV/2,white) named ($\phv{a}_\up{T}$);

        \vertT at (\distH*2,\distV/2,white) named ($\phv{b}_\up{R}$);
        \vertB at (\distH*2,-\distV/2,white) named ($\phv{a}_\up{R}$);

        \vertB at (\distH*2.7,-\distV/2,white) named ($\phv{a}_\up{F}$);

        \node[shift={(\distH/2,0)}, anchor=north] at (0,1*\distV) {\normalsize RF frontend};
        \node[shift={(1.5*\distH,0)}, anchor=north] at (0,1*\distV) {\normalsize tuning network};
        \node[shift={(2.6*\distH,0)}, anchor=north] at (0,1*\distV) {\normalsize radiating structure};

        \draw[shift={(\distH,0)}, line width=1, dashed, dash pattern=on 6pt off 3pt] (0,1*\distV) -- (0,.8*\distV);
        \draw[shift={(2*\distH,0)}, line width=1, dashed, dash pattern=on 6pt off 3pt] (0,1*\distV) -- (0,.8*\distV);
        
    \end{tikzpicture}}

\caption{Signal-flow graph of the model used in this paper. The nodes represent (i) complex vectors corresponding to voltages and circuit-theoretic power waves, and (ii) elements in an $L^2$-space that describe the far-field radiation pattern of the outgoing electromagnetic waves. The edges denote bounded linear operators acting between the respective spaces. 
The original graph shown in~\cite[Fig.~5]{stutz_schwan_studer_efficient_and_physically_consistent_modeling_of_reconfigurable_electromagnetic_structures} additionally contains received voltages, influence from various noise sources, and incoming far-field waves.}

\label{fig:signal_flow_graph}
\end{figure}

To describe the interaction between the radiating structure and electromagnetic waves in its far-field region\footnote{Specifically, our formalism captures the interaction with (i) waves incident from sufficiently far away and (ii) waves radiated into the far-field region.}, we rely on the formalism introduced in~\cite[Sec.~II-B4]{stutz_schwan_studer_efficient_and_physically_consistent_modeling_of_reconfigurable_electromagnetic_structures}. 
To this end, we note that, far away from a radiating structure, the electric field can be approximated using
\begin{align}
    \lim_{r\rightarrow\infty}
    r
    \phv{E}(r,\theta,\varphi)
    &=
    \phv{E}^\swarrow(\theta,\varphi)e^{+jkr}
    \,+\,
    \phv{E}^\nearrow(\theta,\varphi)e^{-jkr},
    \label{eq:far_field_superposition}
\end{align}
where $\phv{E}^\swarrow(\theta,\varphi)$ and $\phv{E}^\nearrow(\theta,\varphi)$ are complex-valued three-dimensional vectors~\cite[Eq.~5.12]{nieto_vesperinas_scattering_and_diffraction_in_physical_optics}. 
Consequently, 
we define the \emph{outgoing far-field power wave pattern} as 
\begin{align}
    \phv{a}_\up{F}:\Omega\rightarrow\mathbb{C}^2,
    \,
    (\theta,\varphi)\mapsto
    \frac{1}{\sqrt{Z_0}}
    \begin{bmatrix}
        \hat{\vect{\theta}}^\T\vect{E}^\nearrow(\theta,\varphi)
        \\
        \hat{\vect{\varphi}}^\T\vect{E}^\nearrow(\theta,\varphi)
    \end{bmatrix}.
    \label{eq:definition_far_field_power_waves_a}
\end{align}
In~\cite{stutz_schwan_studer_efficient_and_physically_consistent_modeling_of_reconfigurable_electromagnetic_structures}, we treat a REMS as a general transceiver and scatterer.  
Because the case study presented in this paper considers a purely transmitting system, our model reduces to the system depicted in~\fref{fig:basic_structure} and described in the following.\footnote{For simplicity, some definitions used here are not fully consistent with~\cite{stutz_schwan_studer_efficient_and_physically_consistent_modeling_of_reconfigurable_electromagnetic_structures}.} 
Our model comprises three subsystems: the \emph{RF frontend}, the \emph{radiating structure}, and the \emph{tuning network}.
The \emph{RF frontend} consists of $N\in\mathbb{N}$ power amplifiers (PAs), which we model using their Thévenin-equivalent circuits. The respective equivalent voltages and impedances are denoted by 
\begin{align}
        \phv{v}_\up{Tx}
        &\hspace{-1pt}\triangleq
        [\phs{v}_{\up{Tx},1}
        \cdots\,
        \phs{v}_{\up{Tx},N}
        ]^\T
    \\
        \mat{Z}_\up{Tx}
        &\hspace{-1pt}\triangleq
        \diag\big([Z_{\up{Tx},1}
        \cdots\,
        Z_{\up{Tx},N}]\big).
\end{align}
The \emph{radiating structure} consists of $M\in\mathbb{N}$ antennas which we model with (i) the \emph{inter-element coupling} \emph{operator}~\mbox{$\oper{S}_{\up{R}_{\up{R}\up{R}}}\hspace{-1mm}:\mathbb{C}^M\rightarrow\mathbb{C}^M$} and (ii) the \emph{transmitting operator}~\mbox{$\oper{S}_{\up{R}_{\up{F}\up{R}}}:\mathbb{C}^M\rightarrow L^2$}.\footnote{To be specific, the $L^2$ space we use here is the Bochner space induced by the measure space $\big( \Omega,\,\mathcal{A},\,\mu \big)$ and the Hilbert space $\mathbb{C}^2$, where $\mathcal{A}$ is the respective $\sigma$-algebra of Lebesgue measurable sets, and $\mu$ is the measure defined by $\mu:\mathcal{A}\rightarrow[0,\infty],\,X\mapsto\iint_{(\theta,\varphi)\in X} \sin(\theta)\, \up{d}\theta\, \up{d}\varphi$.} 
Given the circuit-theoretic power waves~\mbox{$\phv{a}_{\up{R}}\in\mathbb{C}^M$} incident on the $M$ antenna ports, then the circuit-theoretic power waves $\phv{b}_{\up{R}}\in\mathbb{C}^M$ leaving the ports, and the far-field power-wave pattern $\phv{a}_\up{F}\in L^2$ radiated into the far-field region are given as
\begin{align}
    \begin{bmatrix} \phv{b}_{\up{R}} \\ \phv{a}_\up{F} \end{bmatrix} = 
    \begin{bmatrix}
    \oper{S}_{\up{R}_{\up{R}\up{R}}} \\
    \oper{S}_{\up{R}_{\up{F}\up{R}}} 
    \end{bmatrix} \!
    \phv{a}_{\up{R}}.
    \label{eq:matrix_SF}
\end{align}
The \emph{tuning network} contains the (reconfigurable) RF circuitry used for analog beamforming and matching. 
We represent the tuning network as a multiport, with the input-output relation  
\begin{align}
    \begin{bmatrix} \phv{b}_\up{T} \\ \phv{a}_\up{R} \end{bmatrix} = 
    \underbrace{
    \begin{bmatrix}
    \mat{S}_{\up{T}_{\up{T}\up{T}}} & \mat{S}_{\up{T}_{\up{T}\up{R}}} \\
    \mat{S}_{\up{T}_{\up{R}\up{T}}} & \mat{S}_{\up{T}_{\up{R}\up{R}}}
    \end{bmatrix}
    }_{\triangleq \,\mat{S}_\up{T}}
    \begin{bmatrix} \phv{a}_\up{T} \\ \phv{b}_\up{R} \end{bmatrix}\!.
    \label{eq:matrix_ST}
\end{align}
Here, $\phv{a}_\up{T}\in\mathbb{C}^{N}$ and $\phv{b}_\up{T}\in\mathbb{C}^{N}$ represent the incoming and outgoing waves (from the tuning network's perspective) present at the ports connecting the tuning network to the RF frontend. 
\subsection{Input-Output Relationship}
The behavior of the REMSs considered here can be described by a system of linear equations, including~\fref{eq:matrix_SF} and~\fref{eq:matrix_ST}. 
This system of equations can be visualized as the signal-flow graph depicted in~\fref{fig:signal_flow_graph} where we use the auxiliary matrices
\begin{align}
    \mat{K}_{\phv{v}_\up{Tx}}
    &\triangleq
    (\mat{Z}_{\up{Tx}}+R_\up{0}\mat{I}_{N})^{-1}\sqrt{R_\up{0}}
    \\
    \mat{S}_{\up{RF}}
    &\triangleq
    (\mat{Z}_{\up{Tx}}+R_\up{0}\mat{I}_{N})^{-1}(\mat{Z}_{\up{Tx}}-R_\up{0}\mat{I}_{N}).
\end{align}
Utilizing this signal-flow graph offers the advantage that the input-output relationship(s) can be derived quickly.\footnote{The return loop method introduced in~\cite{riegle_lin_matrix_signal_flow_graphs_and_an_optimum_topological_method_for_evaluating_their_gains}, for example, can be used to derive input-output relationships. The return loop method generalizes Mason’s gain formula~\cite{mason_feedback_theory} to nodes in \emph{vector spaces} of arbitrary dimension.}
Since this paper focuses on transmitting REMSs, the REMS input consists only of the PA’s equivalent voltages $\phv{v}_\up{Tx}$, and the only output is the outgoing far-field radiation pattern $\phv{a}_\up{F}$. 
The corresponding input-output relationship is given by 
\begin{align}
    \phv{a}_\up{F}
    &=
    \mathbb{G}_{\phv{v}_\up{Tx}}^{\phv{a}_\up{F}}
    \phv{v}_\up{Tx}.
\end{align}
Here, we use the gain operator
\begin{align}
    \mathbb{G}_{\phv{v}_\up{Tx}}^{\phv{a}_\up{F}}
    &\triangleq
    \oper{S}_{\up{R}_\up{FR}}
    (\mat{I}_M-\mat{L}_\up{2})^{-1}
    \mat{S}_{\up{T}_\up{RT}}
    (\mat{I}_N-\mat{L}_\up{1}-\mat{L}_\up{3})^{-1}
    \mat{K}_{\phv{v}_\up{Tx}},
\end{align}
where we use the following auxiliary matrices:
\begin{align}
    \mat{L}_\up{1}
    &\triangleq
    \mat{S}_\up{RF}
    \mat{S}_{\up{T}_\up{TT}}
    \\
    \mat{L}_\up{2}
    &\triangleq
    \mat{S}_{\up{T}_\up{RR}}
    \oper{S}_{\up{R}_\up{RR}}
    \\
    \mat{L}_\up{3}
    &\triangleq
    \mat{S}_\up{RF}
    \mat{S}_{\up{T}_\up{TR}}
    \oper{S}_{\up{R}_\up{RR}}
    \left(
    \mat{I}_{M}
    -\mat{L}_\up{2}
    \right)^{-1}
    \mat{S}_{\up{T}_\up{RT}}.
\end{align}

\subsection{Power Metrics}
In~\cite[Def.~5]{stutz_schwan_studer_efficient_and_physically_consistent_modeling_of_reconfigurable_electromagnetic_structures}, we define the following \emph{interface-related power metrics} to analyze the power flow between a REMS's subsystems: The \emph{tuning network accepted power} is\footnote{For definitions of the voltage and current vectors $\phv{v}_\up{T}$, $\phv{i}_\up{T}$, $\phv{v}_\up{R}$, and $\phv{i}_\up{R}$, we refer the reader to \fref{fig:basic_structure} or~\cite{stutz_schwan_studer_efficient_and_physically_consistent_modeling_of_reconfigurable_electromagnetic_structures}.
}
\begin{align}
    \label{eq:defi_PT}
    \hspace{.5cm}P_\up{T}
    &\triangleq
    \mathmakebox[2cm][c]{\;\Re\{\phv{v}_\up{T}^\He\phv{i}_\up{T}\}}
    =
    \|\phv{a}_\up{T}\|_2^2
    -
    \|\phv{b}_\up{T}\|_2^2,
    \intertext{
    the \emph{radiating structure accepted power} is 
    }
    \label{eq:defi_PR}
    P_\up{R}
    &\triangleq
    \mathmakebox[2cm][c]{\;
    \Re\{\phv{v}_\up{R}^\He\phv{i}_\up{R}\}
    }
    =
    \|\phv{a}_\up{R}\|_2^2
    -
    \|\phv{b}_\up{R}\|_2^2,
    \intertext{
    and the \emph{total far field radiated power} of a pure transmitter is  
    }
    \label{eq:defi_PF}
    P_\up{F}
    &=
    \mathmakebox[2cm][c]{\;
    P^\nearrow 
    }
    =
    \|\phv{a}_\up{F}\|_{L^2}^2.
    &
\end{align}
Here, with $P^\nearrow$ we denote the power radiated by the radiating structure in all directions.

Moreover, in~\cite[Def.~6]{stutz_schwan_studer_efficient_and_physically_consistent_modeling_of_reconfigurable_electromagnetic_structures}, 
we define the \emph{power amplifiers available power} to analyze the potentially available power flow out of the PAs: 
For a fixed PA output voltage vector $\phv{v}_\up{Tx}\in\mathbb{C}^{N}$, the \emph{power amplifiers available power} is given as
    \begin{align}
        \label{eq:defi_PA}
        P_\up{A}
        =
        \frac{1}{4}\phv{v}_\up{Tx}^\He
        \Re\{\mat{Z}_\up{Tx}\}^{-1}
        \phv{v}_\up{Tx}.
    \end{align}

\subsection{Radiating Intensity and Gain of a REMS}
We use the \emph{radiation intensity} $I(\theta,\varphi)\triangleq\|\phv{a}_\up{F}(\theta,\varphi)\|_2^2$ to determine how much power is transmitted per unit solid angle into the direction $(\theta,\varphi)$. 
Furthermore, we use the \emph{REMS gain}, which, for a given PA output voltage vector~$\phv{v}_\up{Tx}\in\mathbb{C}^{N}$, and a given direction~\mbox{$(\theta,\varphi)\in\Omega$} is given as 
%
\begin{align}
    G_\up{REMS}(\phv{v}_\up{Tx};\,\theta,\varphi)
    & = 
    \frac{1}{P_\up{A}}
    4\pi 
    I(\theta,\varphi)
    \left\rvert{
        {
        \scriptscriptstyle
        \begin{subarray}{l}
            \hspace{1mm}        \\
            \hspace{1mm}        \\
            \hspace{1mm}        \\
            \phv{v}_\up{Tx} \up{ fixed}   
        \end{subarray}
        }
    }
    \right.
    \\
    &  =
    16\pi
    \frac{
    \|
    \big(
        \mathbb{G}_{\phv{v}_\up{Tx}}^{\phv{a}_\up{F}}
        \phv{v}_\up{Tx}
    \big)
    (\theta,\varphi)
    \|_2^2}
    {\phv{v}_\up{Tx}^\He
    \Re\{\mat{Z}_\up{Tx}\}^{-1}
    \phv{v}_\up{Tx}}.
    \label{eq:defi_G_REMS}
\end{align}
As shown in~\fref{fig:power_flow_graph}, the \emph{REMS gain} characterizes the combined effect of a general REMS's (i) \emph{matching efficiency}~\mbox{$\eta_\up{matching}\triangleq P_\up{T}/P_\up{A}$}, (ii) \emph{tuning network efficiency} \mbox{$\eta_\up{tuning}\triangleq P_\up{R}/P_\up{T}$}, (iii) \emph{radiation efficiency} \mbox{$\eta_\up{radiating}\triangleq P_\up{F}/P_\up{R}$}, and (iv) directivity \mbox{$D(\theta,\varphi)\triangleq (4\pi I(\theta,\varphi))/P_\up{F}$}. 
\begin{rem}[cf.~{[}1, Rem.~6{]}]
    The REMS gain~$G_\up{REMS}$ can be interpreted as the radiation intensity gain of the REMS relative to a reference system containing a single lossless, isotropic radiator\footnote{A coherent isotropic radiator cannot exist~\cite{mathis_a_short_proof_that_an_isotropic_antenna_is_impossible}.}, whose feed is perfectly matched to the reference system PA's output stage provided that~$P_\up{A}$ of the REMS is equal to~$P_\up{A}$ of the reference system. 
\end{rem}

To complement the REMS gain, we now introduce two additional gain metrics: the \emph{tuning gain} and the \emph{radiating gain}. 

\begin{defi}[Tuning Gain]\label{defi:tuning_transmitting_gain}
    Given a tuning network and a radiating structure. 
    Let $\phv{a}_\up{T}\in\mathbb{C}^{N}$ be the power waves flowing into the tuning network and $(\theta,\varphi)\in\Omega$ specify a direction. 
    We define the \emph{tuning gain} corresponding to $\phv{a}_\up{T}$ and $(\theta,\varphi)$ as 
    \begin{align}
        G_\up{T}(\phv{a}_\up{T};\,\theta,\varphi)
        & \triangleq
        \frac{1}{P_\up{T}}
        4\pi 
        I(\theta,\varphi)
        \left\rvert{
            {
            \scriptscriptstyle
            \begin{subarray}{l}
                \hspace{1mm}        \\
                \hspace{1mm}        \\
                \hspace{1mm}        \\
                \phv{a}_\up{T} \up{ fixed}   
            \end{subarray}
            }
        }
        \right.
        \\
        & =
        4\pi
        \frac{
        \|
        \big(
            \mathbb{G}_{\phv{a}_\up{T}}^{\phv{a}_\up{F}}
            \phv{a}_\up{T}
        \big)
        (\theta,\varphi)
        \|_2^2}
        {
        \|
            \phv{a}_\up{T}
        \|_2^2
        -
        \|
        \mathbb{G}_{\phv{a}_\up{T}}^{\phv{b}_\up{T}}
        \phv{a}_\up{T}
        \|_2^2}
        ,
        \label{eq:defi_G_T}
    \end{align}
    with 
    \begin{align}
        \mathbb{G}_{\phv{a}_\up{T}}^{\phv{a}_\up{F}}
        &\triangleq 
        \oper{S}_{\up{R}_\up{FR}}
        (\mat{I}_M-\mat{L}_\up{2})^{-1}
        \mat{S}_{\up{T}_\up{RT}}
        \\
        \mathbb{G}_{\phv{a}_\up{T}}^{\phv{b}_\up{T}}
        &\triangleq 
        \mat{S}_{\up{T}_\up{TT}}
        +
        \mat{S}_{\up{T}_\up{TR}}
        \oper{S}_{\up{R}_\up{RR}}
        (\mat{I}_M-\mat{L}_\up{2})^{-1}
        \mat{S}_{\up{T}_\up{RT}}.
    \end{align}
\end{defi}

\begin{defi}[Radiating Gain]\label{defi:radiation_transmitting_gain}
    Given a radiating structure.  
    Let $\phv{a}_\up{R}\in\mathbb{C}^{M}$ be the power waves flowing into the radiation structure and $(\theta,\varphi)\in\Omega$ specify a direction. 
    We define the \emph{radiating gain} corresponding to $\phv{a}_\up{R}$ and $(\theta,\varphi)$ as 
    \begin{align}
        G_\up{R}(\phv{a}_\up{R};\,\theta,\varphi)
        & \triangleq
        \frac{1}{P_\up{R}}
        4\pi 
        I(\theta,\varphi)
        \left\rvert{
            {
            \scriptscriptstyle
            \begin{subarray}{l}
                \hspace{1mm}        \\
                \hspace{1mm}        \\
                \hspace{1mm}        \\
                \phv{a}_\up{R} \up{ fixed}   
            \end{subarray}
            }
        }
        \right.
        \\
        & =
        4\pi
        \frac{
        \|
        \big(
            \mathbb{G}_{\phv{a}_\up{R}}^{\phv{a}_\up{F}}
            \phv{a}_\up{R}
        \big)
        (\theta,\varphi)
        \|_2^2}
        {
        \|
            \phv{a}_\up{R}
        \|_2^2
        -
        \|
        \mathbb{G}_{\phv{a}_\up{R}}^{\phv{b}_\up{R}}
        \phv{a}_\up{R}
        \|_2^2}
        ,
        \label{eq:defi_G_R}
    \end{align}
    with 
    \begin{align}
        \mathbb{G}_{\phv{a}_\up{R}}^{\phv{a}_\up{F}}
        \triangleq 
        \oper{S}_{\up{R}_\up{FR}} \quad \text{and} \quad
        \mathbb{G}_{\phv{a}_\up{R}}^{\phv{b}_\up{R}}
        \triangleq 
        \oper{S}_{\up{R}_\up{RR}}.
    \end{align}
\end{defi}
\def\powerNode at (#1,#2,#3){
    \begin{scope}[shift={(#1, #2)}]
        \draw  [line width=1pt, fill=white] (0,0) ellipse (0.4 and 0.4);
        \node[color=black] at (0,0) {#3};     
    \end{scope}
}

\def\powerNodeII at (#1,#2,#3){
    \begin{scope}[shift={(#1, #2)}]
        \draw[rounded corners = 11.5, line width=1pt, fill=white] (-.8, -.4) rectangle (.8, .4) {};
        \node[color=black] at (0,0) {#3};     
    \end{scope}
}

\def\causalRelation at (#1,#2,#3,#4,#5,#6,#7){

    \draw [line width=1pt, color=blue_120, arrows = {-Stealth[inset=0, length=6pt, angle'=35]}] (#1,#2) -- (#3*0.55+#1*0.45,#4*0.55+#2*0.45);
    \draw [line width=1pt, color=blue_120] (#3,#4) -- (#3/2+#1/2,#4/2+#2/2);

    \node[shift={(#3*0.5+#1*0.5,#4*0.5+#2*0.5)}, anchor=north] at (#6,#7) {\color{blue_120}#5};
} 

\def\gainArrow at (#1,#2,#3,#4,#5,#6,#7,#8){
    \draw [densely dotted, line width=1pt, color=green, arrows = {-Stealth[inset=0, length=6pt, angle'=35]}] (2*\distPowerNodes,#8) -- (2.1*\distPowerNodes,#8);

    \draw [densely dashed, line width=1pt, color=green] plot[smooth, tension=.8] coordinates {(0,0) (2*\distPowerNodes,#8) (4.2*\distPowerNodes,0)};
  
    \node[shift={(2.1*\distPowerNodes,#8)}, anchor=south] at (#6,#7) {\color{green}#5};
}

\def\gainArrowT at (#1,#2,#3,#4,#5,#6,#7,#8){
    \draw [densely dotted, line width=1pt, color=green, arrows = {-Stealth[inset=0, length=6pt, angle'=35]}] (2.5*\distPowerNodes,#8) -- (2.6*\distPowerNodes,#8);

    \draw [densely dashed, line width=1pt, color=green] plot[smooth, tension=.8] coordinates {(1*\distPowerNodes,0) (2.5*\distPowerNodes,#8) (4.2*\distPowerNodes,0)};
  
    \node[shift={(2.5*\distPowerNodes,#8)}, anchor=south] at (#6,#7) {\color{green}#5};
}

\def\gainArrowR at (#1,#2,#3,#4,#5,#6,#7,#8){
    \draw [densely dotted, line width=1pt, color=green, arrows = {-Stealth[inset=0, length=6pt, angle'=35]}] (3*\distPowerNodes,#8) -- (3.1*\distPowerNodes,#8);

    \draw [densely dashed, line width=1pt, color=green] plot[smooth, tension=.8] coordinates {(2*\distPowerNodes,0) (3*\distPowerNodes,#8) (4.2*\distPowerNodes,0)};
  
    \node[shift={(3*\distPowerNodes,#8)}, anchor=south] at (#6,#7) {\color{green}#5};
}

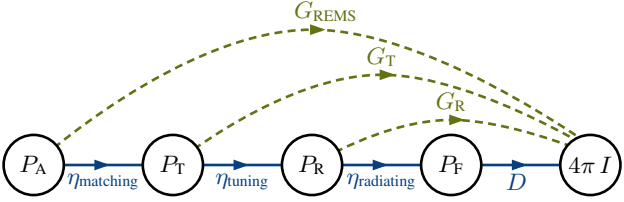
\begin{figure}[btp]
    \vspace{-.25cm} 
    \centering
    {
    \small
    \begin{tikzpicture}
        \def\distPowerNodes{1.85};

        \causalRelation at (0,0,\distPowerNodes,0,$\eta_\up{matching}$,0,0);
        \causalRelation at (\distPowerNodes,0,2*\distPowerNodes,0, $\eta_\up{tuning}$ ,0,0);
        \causalRelation at (2*\distPowerNodes,0,3*\distPowerNodes,0, $\eta_\up{radiating}$ ,0,0);
        \causalRelation at (3*\distPowerNodes,0,4*\distPowerNodes,0, {$D$ } ,0,0);

        \gainArrow at (0, 0, 3*\distPowerNodes, 0.8*\distPowerNodes, {$G_\up{REMS}$}, 0, 0, 1.8);
        
        \gainArrowT at (0, 0, 3*\distPowerNodes, 0.8*\distPowerNodes, {$G_\up{T}$}, 0, 0, 1.2);
                
        \gainArrowR at (0, 0, 3*\distPowerNodes, 0.8*\distPowerNodes, {$G_\up{R}$}, 0, 0, 0.6);
       
        \powerNode at (0,0,$P_\up{A}$);
        \powerNode at (\distPowerNodes, 0,$P_\up{T}$);
        \powerNode at (2*\distPowerNodes, 0,$P_\up{R}$);
        \powerNode at (3*\distPowerNodes, 0,$P_\up{F}$);
        \powerNode at (4*\distPowerNodes, 0,{$4\pi \,I$});

    \end{tikzpicture}}

\caption{
The relationships between the power metrics, including the radiating intensity, are illustrated with solid blue lines. The REMS gain, tuning gain, and radiating gain, which relate the radiation intensity to the corresponding power quantities, are shown with a dashed green line. 
This figure extends~\cite[Fig.~6]{stutz_schwan_studer_efficient_and_physically_consistent_modeling_of_reconfigurable_electromagnetic_structures} by additionally including the tuning and radiating gain.
}

\label{fig:power_flow_graph}
\end{figure}
\begin{figure}[!t]
    \centering
    \vspace{-1cm}
    \subfloat[$4 \times 4$]{
        \centering
        \includegraphics[width=0.48\columnwidth,trim={1500 500 1500 500}, clip]{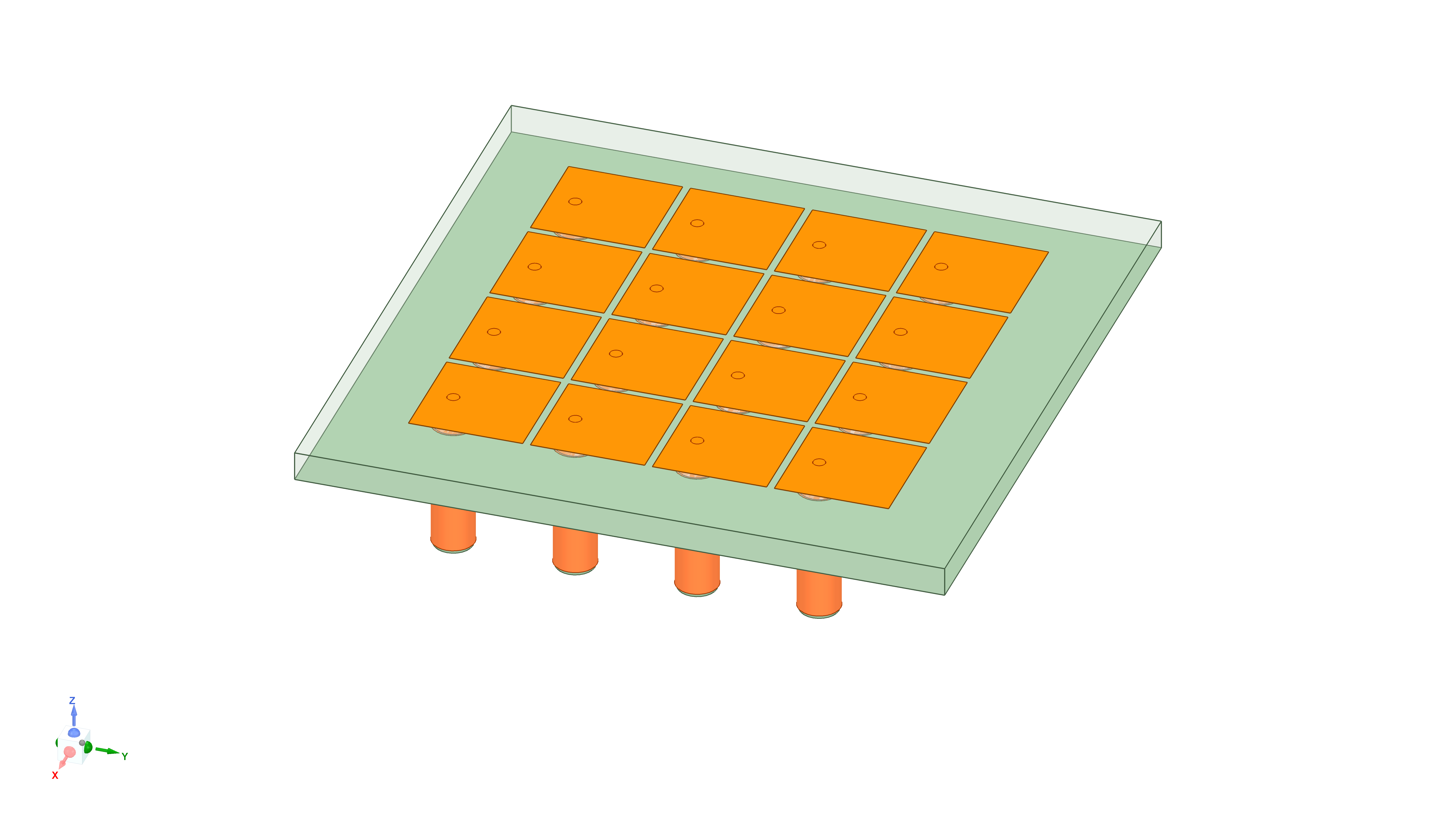}
        \label{fig:radiating_structure_4}
    }
    \subfloat[$16 \times 16$]{
        \centering
        \includegraphics[width=0.48\columnwidth, trim={1500 200 1500 300}, clip]{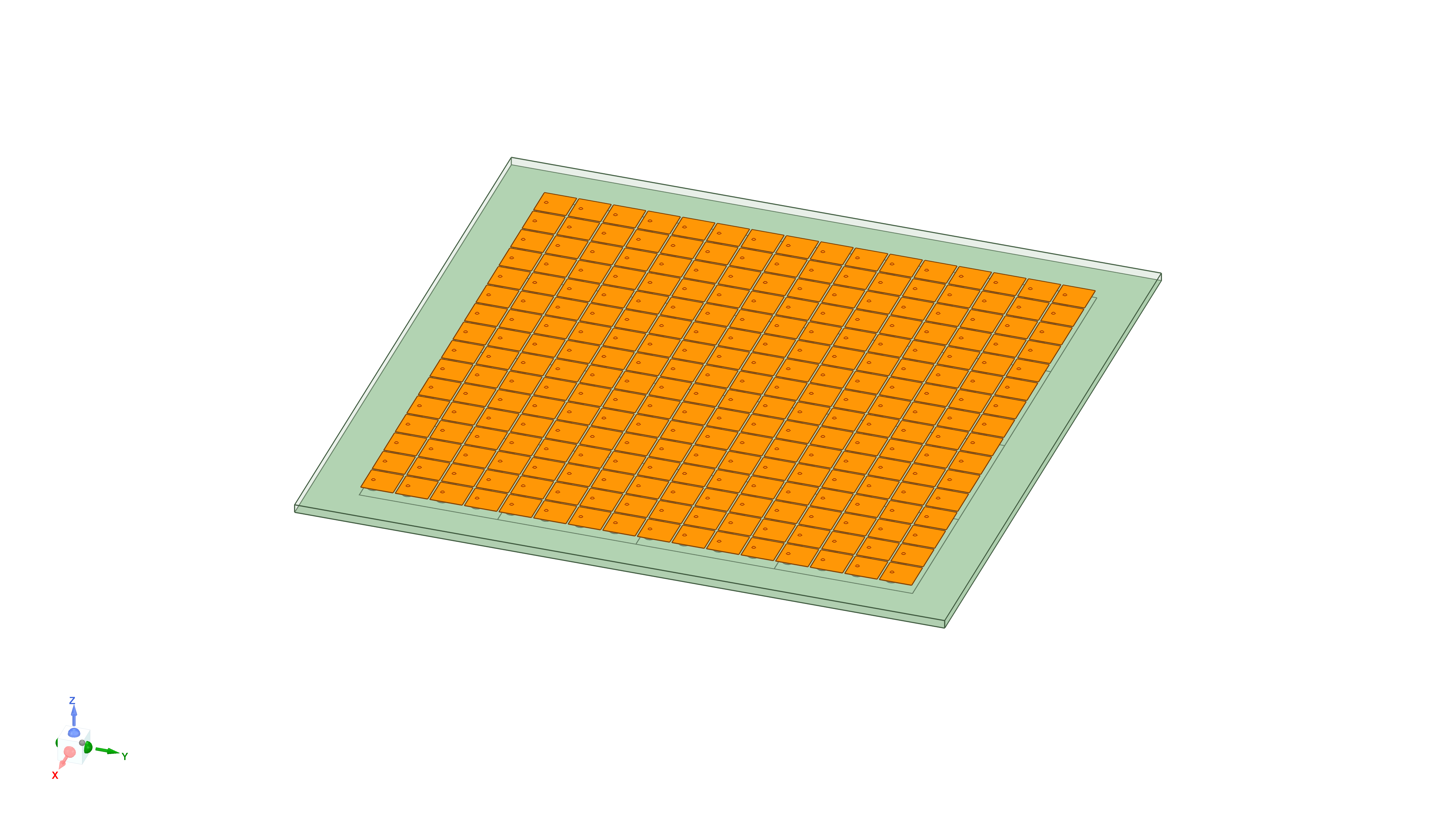}
        \label{fig:radiating_structure_16}
    }
    \caption{
    Screenshots from Ansys HFSS of the used radiating structures: A $4\times4$ and a $16\times16$ ultra dense arrays of square patch antennas with an inter-element spacing of $\frac{\lambda}{4}$. 
    The arrays lie on the $\theta = 90^\circ$ plane.}
    \label{fig:radiating_structure}
\end{figure}

\section{A Low-Complexity Beamforming Architecture}\label{sec:case_study}
We now demonstrate that with \emph{physically consistent} modeling, one can jointly perform hybrid beamforming and matching in an ultra-dense massive antenna array using inexpensive RF circuitry. 
Concretely, given the radiating structure detailed next, we propose a \emph{cost-efficient}\footnote{We use \emph{cost-efficient} to emphasize that our hybrid beamformer needs only a fraction of the RF chains of an \emph{all-digital} architecture and relies on low-cost RF switches rather than expensive phase shifters for analog beamforming.} beamsteering architecture. 
\subsection{Radiation Structure}\label{sec:used_radiating_structure}

We consider the $4 \times 4$ and $16 \times 16$ antenna arrays depicted  in~\fref{fig:radiating_structure} operating at \SI{12}{\giga \hertz}.
We use square patch antennas with side lengths of approximately $\frac{\lambda}{4}$. 
The antennas are excited by a coaxial port, and the input impedance of a single element, when embedded in the array, is roughly \SI{50}{\ohm}.\footnote{The exact input impedances vary due to mutual coupling.} 
In both arrays, the inter-element spacing between element centers is $\frac{\lambda}{4}$; hence, they are strongly coupled, ultra-dense arrays.
We characterized both arrays, specifically their \emph{inter-element coupling operator} $\oper{S}_{\up{R}_\up{RR}}$ and \emph{transmitting operator} $\oper{S}_{\up{R}_\up{FR}}$, using full-wave EM simulations in Ansys HFSS. 
\begin{figure}[!t]
    \vspace{-.75cm}
    \centering
    \hfill
    \subfloat[$4 \times 4$]{
        \centering
        \includegraphics{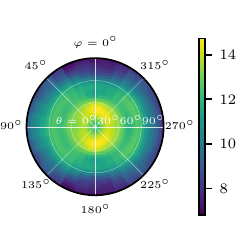}
        \label{fig:radiation_gain_4}
    }
    \subfloat[$16 \times 16$]{
        \centering
        \includegraphics{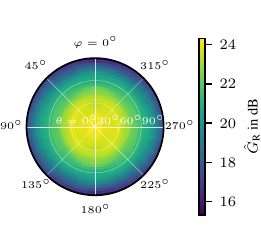}
        \label{fig:radiation_gain_16}
    }
    \caption{Radiating structure gain $\hat{G}_\up{R}(\theta,\varphi)$ of the used radiating structures for $(\theta,\varphi) \in [0, \pi/2] \times [0, 2\pi)$.}
    \label{fig:radiation_gain}
\end{figure}

\subsection{Gain Metrics}  
The \emph{REMS gain} $G_\up{REMS}(\phv{v}_\up{Tx};\theta,\varphi)$ defined in~\cite[Def.~7]{stutz_schwan_studer_efficient_and_physically_consistent_modeling_of_reconfigurable_electromagnetic_structures} provides a comprehensive metric for evaluating hybrid beamformers, as it accounts for directivity, ohmic losses, impedance mismatch, and the effects of both analog and digital beamforming. 
Two further gain metrics useful for assessing individual subsystems of a REMS are the \emph{tuning gain} $G_\up{T}(\phv{a}_\up{T};\theta,\varphi)$ (see~\fref{defi:tuning_transmitting_gain}) and the \emph{radiating gain} $G_\up{R}(\phv{a}_\up{R};\theta,\varphi)$ (see~\fref{defi:radiation_transmitting_gain}). 
All three gain metrics depend not only on the direction~$(\theta,\varphi)$ but also on the current beamforming scheme, via~$\phv{v}_\up{Tx}$, $\phv{a}_\up{T}$, and~$\phv{a}_\up{R}$. 
In the following, when evaluating a gain metric in a given direction $(\theta,\varphi)$, we choose~\mbox{$\phv{v}_\up{Tx}\in\mathbb{C}^N$},~\mbox{$\phv{a}_\up{T}\in\mathbb{C}^N$}, and $\phv{a}_\up{R}\in\mathbb{C}^M$ so as to maximize~$G_\up{REMS}(\phv{v}_\up{Tx};\theta,\varphi)$, $G_\up{T}(\phv{a}_\up{T};\theta,\varphi)$, and $G_\up{R}(\phv{a}_\up{R};\theta,\varphi)$, respectively. 
We now show how to obtain these voltage and power wave vectors.

It follows directly from~\fref{eq:defi_G_REMS}, \fref{eq:defi_G_T}, and~\fref{eq:defi_G_R} that, in all three cases, the respective maximization problem can be expressed as
\begin{align}
\label{eq:Rayleigh_quotient}
    \argmax_{\phv{x}}
    G(\phv{x};\theta,\varphi) = \frac{\phv{x}^\He\mat{A}(\theta,\varphi)\phv{x}}{\phv{x}^\He\mat{B}\phv{x}},
\end{align} 
where $\phv{x}$ denotes the voltage or power-wave vector being optimized, and $\mat{A}(\theta,\varphi)$ and $\mat{B}$ are Hermitian complex matrices chosen such that $\phv{x}^\He \mat{A}(\theta,\varphi)\phv{x}=4\pi I(\theta,\varphi)$ and $\phv{x}^\He \mat{B}\phv{x}$ equals a power metric.\footnote{Explicitly, for a fixed direction $(\theta,\varphi)\in\Omega$, these matrices are given as follows: 
When maximizing $G_\up{REMS}$, we use $\mat{A}=4\pi(\oper{G}_{\phv{v}_\up{Tx}}^{\phv{a}_\up{F}})^\He\oper{G}_{\phv{v}_\up{Tx}}^{\phv{a}_\up{F}}$ and $\mat{B}=1/4\Re\{\mat{Z}_\up{Tx}\}^{-1}$; 
when maximizing $G_\up{T}$, we use $\mat{A}=4\pi(\oper{G}_{\phv{a}_\up{T}}^{\phv{a}_\up{F}})^\He\oper{G}_{\phv{a}_\up{T}}^{\phv{a}_\up{F}}$ and $\mat{B}=\mat{I}_N-(\oper{G}_{\phv{a}_\up{T}}^{\phv{b}_\up{T}})^\He\oper{G}_{\phv{a}_\up{T}}^{\phv{b}_\up{T}}$; 
and when maximizing $G_\up{R}$, we use $\mat{A}=4\pi(\oper{G}_{\phv{a}_\up{R}}^{\phv{a}_\up{F}})^\He\oper{G}_{\phv{a}_\up{R}}^{\phv{a}_\up{F}}$ and $\mat{B}=\mat{I}_M-(\oper{G}_{\phv{a}_\up{R}}^{\phv{b}_\up{R}})^\He\oper{G}_{\phv{a}_\up{R}}^{\phv{b}_\up{R}}$. Note that, to simplify notation, we refrain from explicitly indicating the dependence on the direction $(\theta,\varphi)$ in this footnote.  
}
We note that the radiation intensity $I(\theta,\varphi)$ is nonnegative by definition, and we assume the power metrics~$P_\up{A}$,~$P_\up{T}$, and~$P_\up{R}$ are strictly positive.
From $I(\theta,\varphi)\geq0$, it follows that $\mat{A}(\theta,\varphi)$ is positive semidefinite and from~\mbox{$P_\up{A},P_\up{T},P_\up{R}>0$} it follows that $\mat{B}$ is positive definite. 
Consequently,~\fref{eq:Rayleigh_quotient} is a Rayleigh-quotient maximization whose solution is well known~\cite{li_rayleigh_quotient_based_optimization}:
\begin{align}
    \label{eq:solution_optimization_problem}
    \hat{\phv{x}}(\theta,\varphi)\triangleq
    \mat{B}^{-\frac{1}{2}}\vect{c}_1(\theta,\varphi)
    \in
    \argmax_{\phv{x}}
    G(\phv{x};\theta,\varphi).
\end{align}
Here, the matrix $\mat{B}^{-\frac{1}{2}}$ is a positive definite square root of~$\mat{B}^{-1}$ and $\vect{c}_1(\theta,\varphi)$ is the dominant eigenvector of~\mbox{$\mat{C}(\theta,\varphi)\triangleq \mat{B}^{-\frac{1}{2}} \mat{A}(\theta,\varphi) \mat{B}^{-\frac{1}{2}}$}.
Let $\hat{\phv{v}}_\up{Tx}(\theta,\varphi)$, $\hat{\phv{a}}_\up{T}(\theta,\varphi)$, and $\hat{\phv{a}}_\up{R}(\theta,\varphi)$ be optimal voltage and power wave vectors according to~\eqref{eq:solution_optimization_problem}. 
For the rest of this paper, we use the following shorthand notation:
\begin{align}
    \label{eq:simple_G_REMS}
    \hat{G}_\up{REMS}(\theta,\varphi)
    &\triangleq 
    G_\up{REMS}(\hat{\phv{v}}_\up{Tx}(\theta,\varphi);\theta,\varphi)
    \\
    \label{eq:simple_G_T}
    \hat{G}_\up{T}(\theta,\varphi)
    &\triangleq 
    G_\up{T}(\hat{\phv{a}}_\up{T}(\theta,\varphi);\theta,\varphi)
    \\
    \label{eq:simple_G_R}
    \hat{G}_\up{R}(\theta,\varphi)
    &\triangleq 
    G_\up{R}(\hat{\phv{a}}_\up{R}(\theta,\varphi);\theta,\varphi).
\end{align}

\begin{rem}\label{rem:gain_inequality}
    By definition, for every REMS, it holds that
    \begin{align}
        \hat{G}_\up{REMS}(\theta,\varphi)
        \leq 
        \hat{G}_\up{T}(\theta,\varphi)
        \leq
        \hat{G}_\up{R}(\theta,\varphi)
        \hspace{.5cm}
        \forall (\theta,\varphi)\in\Omega.
    \end{align}
\end{rem}
\begin{rem}\label{rem:G_R_explain}
    For a given radiating structure, the \emph{radiating gain}~\mbox{$\hat{G}_\up{R}(\theta,\varphi)$} quantifies the fundamentally maximal REMS gain~\mbox{$\hat{G}_\up{REMS}(\theta,\varphi)$} that can be achieved with that radiating structure. 
    Consequently, the deviation between $\hat{G}_\up{R}(\theta,\varphi)$ and~\mbox{$\hat{G}_\up{REMS}(\theta,\varphi)$} quantifies losses due to non-ideal matching and due to a non-ideal tuning network.\footnote{Note that an ideal tuning network ensures optimal beamforming, regardless of the number of RF chains used.}
\end{rem}
\begin{rem}\label{rem:G_T_explain}
    For a given tuning network and radiating structure, the \emph{tuning gain} $\hat{G}_\up{T}(\theta,\varphi)$ quantifies the fundamentally maximal REMS gain $\hat{G}_\up{REMS}(\theta,\varphi)$ that can be achieved with that tuning network and radiating structure.
    The deviation between~$\hat{G}_\up{T}(\theta,\varphi)$ and $\hat{G}_\up{REMS}(\theta,\varphi)$ quantifies the losses due to non-ideal matching.
\end{rem}
\fref{fig:radiation_gain} shows the radiating gain $\hat{G}_\up{R}(\theta,\varphi)$ for the two used radiating structures.

\subsection{Objective}\label{sec:case_study_objective}
Given the radiating structures described in~\fref{sec:used_radiating_structure}, our objective is to achieve a high REMS gain $\hat{G}_\up{REMS}(\theta,\varphi)$  using a \emph{cost-efficient} architecture for all directions in front of the arrays (i.e., we consider $(\theta,\varphi)\in[0,\pi/2]\times[0,2\pi)$, with the arrays lying in the $\theta=90^\circ$ plane).

\subsection{Proposed Hybrid Beamformer Architecture}\label{sec:proposed_architecture}

\def\ground (#1,#2) {
    \draw[shift={(#1,#2)}, line width=.6] (-0.16,0) -- (0.16,0);
    \draw[shift={(#1,#2)}, line width=.6] (-0.12,-0.05) -- (0.12,-0.05);
    \draw[shift={(#1,#2)}, line width=.6] (-0.08,-0.1) -- (0.08,-0.1);
}

\def\impedance (#1,#2,#3) {
    \draw  [shift={(#1,#2)}, line width=.7pt, anchor=center, fill=white] (-.3,-.12) rectangle (.3,.12);
    \node[shift={(#1,#2)}, anchor=south] at (0,0.12) {#3};
}

\def\voltage (#1,#2) {
    \draw[shift={(#1,#2)}, line width=.7pt] (0,0) ellipse (0.28 and 0.28);
    \draw[shift={(#1,#2)}, line width=.7pt] (0,-.28) -- (0,.28);
}

\def\cont (#1,#2,#3) {
    \draw  [shift={(#1,#2)}, line width=.5pt, fill=grey, color=grey] (-.2,.14)  ellipse (0.02 and 0.02);
    \draw  [shift={(#1,#2)}, line width=.5pt, fill=grey, color=grey] (-.2,0) ellipse (0.02 and 0.02);
    \draw  [shift={(#1,#2)}, line width=.5pt, fill=grey, color=grey] (-.2,-0.14) ellipse (0.02 and 0.02);
    \node[shift={(#1,#2)}, anchor=west] at (-.2,0) {\color{grey} \footnotesize #3};
}

\def\antenna (#1,#2) {
    \draw[shift={(#1,#2)}, line width=.6] (0,0) -- (0,0.4) -- (0.15,0.4) -- (0,0.1) -- (-0.15,0.4) -- (0,0.4);
}

\def\transmission (#1,#2,#3) {
    \draw  [shift={(#1,#2)}, line width=.7] (0,0) ellipse (0.08 and 0.12);
    \draw  [shift={(#1,#2)}, line width=.7, fill=white] (0.6,0) ellipse (0.08 and 0.12);
    \draw[shift={(#1,#2)}, line width=.7] (0,0.12) -- (0.6,0.12);
    \draw[shift={(#1,#2)}, line width=.7] (0,-0.12) -- (0.6,-0.12);
    \node[shift={(#1,#2)}, anchor=south] at (0.3,0.12) {#3};
}

\def\transmissiony (#1,#2,#3) {
    \draw  [shift={(#1,#2)}, line width=.7] (0,0) ellipse (0.12 and 0.08);
    \draw  [shift={(#1,#2)}, line width=.7, fill=white] (0,-0.6) ellipse (0.12 and 0.08);
    \draw[shift={(#1,#2)}, line width=.7] (0.12,-0) -- (0.12,-0.6);
    \draw[shift={(#1,#2)}, line width=.7] (-0.12,0) -- (-0.12,-0.6);
    \node[shift={(#1,#2)}, anchor=west, fill=white] at (0.12,-0.3) {#3};
}

\def\switch (#1,#2) {
    \draw[shift={(#1,#2)}, line width=.6] (0,0) -- (0.3, 0.15);
    \draw[shift={(#1,#2)}, line width=.6] (0.3,0) -- (0.4,0);
}

\def\switchy (#1,#2) {
    \draw[shift={(#1,#2)}, line width=.6] (0,0) -- (0.15, -0.3);
    \draw[shift={(#1,#2)}, line width=.6] (0,-0.3) -- (0,-0.4);
}

\begin{figure*}[btp]
    \centering
    \subfloat[proposed network architecture for one tile]{
    \centering
    {
    \small
    \begin{tikzpicture}
        \def\height{1};
        
        \begin{scope}[shift={(-0.15,0)}]
            \draw[line width=.6] (4.6,0) -- (5.2,0);
            \draw[line width=.6] (5.8,0) -- (6.1,0);

            \draw[line width=.6] (5.8,0) -- (6.1,0);
            \draw[black, fill=black] (6.1,0) circle (0.04);
            
            \draw[line width=.6] (6.1,\height) -- (6.1,-\height);

            \draw[line width=.6] (6.1,\height) -- (6.4,\height);
            \draw[line width=.6] (7.0,\height) -- (7.3,\height);
            \draw[line width=.6] (8.1,\height) -- (8.9,\height);
            \antenna (8.9, \height)
            
            \draw[line width=.6] (6.1,-\height) -- (6.4,-\height);
            \draw[line width=.6] (7.0,-\height) -- (7.3,-\height);
            \draw[line width=.6] (8.1,-\height) -- (8.9,-\height);
            \antenna (8.9,-\height)

            \transmission (5.2,0, $\frac{50}{16}\,\Omega$)
            
            \transmission (6.4, \height, $50\,\Omega$)
            \transmission (6.4, -\height, $50\,\Omega$)
            
            \draw  [shift={(7.9,\height)}, line width=.7pt, anchor=center, fill=white] (-0.6,-0.5) rectangle (0.6,0.5);
            \node[shift={(7.9,\height)}, anchor=center, align=center] at (0,0) {switch\\unit};

            \draw  [shift={(7.9,-\height)}, line width=.7pt, anchor=center, fill=white] (-0.6,-0.5) rectangle (0.6,0.5);
            \node[shift={(7.9,-\height)}, anchor=center, align=center] at (0,0) {switch\\unit};
    
            \cont (7.5,0, $(16)$)
         
        \end{scope}

        \draw  [line width=.7pt, anchor=center, fill=white] (3.2,-0.7) rectangle (4.8,0.7);
        \node[anchor=center, align=center] at (4,0) {proposed\\matching\\network};

        \begin{scope}[shift={(0,0)}]
    
            \draw[line width=.6] (2,0) -- (3.2,0);
    
            \draw[line width=.6] (2,-0.9) -- (2,0);
            
            \impedance (2.7,0, $\frac{50}{16}\,\Omega$)
    
            \voltage (2,-0.45)
    
            \ground (2,-0.9)
        \end{scope}

    \begin{scope}[shift={(0,0)}]
    \end{scope}
        
    \end{tikzpicture}}
    \label{fig:proposed_tuning_network}
    }
    \hfill
    \subfloat[proposed matching network]{
        \centering
        {
        \small
        \begin{tikzpicture}
            \def\height{0.8};
            
            \def\mcx{0};
            \def\mcy{0};
    
            \draw  [line width=.5pt, anchor=center, dashed, color=grey] (-1.7,-2.3) rectangle (1.7,0.9);

            \draw[line width=.6] (\mcx-0.3,\mcy) -- (\mcx+0.3,\mcy);
            
            \draw[line width=.6] (\mcx-1.9,\mcy) -- (\mcx-0.9,\mcy);
            \draw[line width=.6] (\mcx+0.9,\mcy) -- (\mcx+1.9,\mcy);
            
            \transmission (\mcx+0.3,\mcy, $\frac{1}{8}\,\lambda$)
            \transmission (\mcx-0.9,\mcy, $\frac{1}{8}\,\lambda$)
    
            \foreach \x in {-1.2,0,1.2} {
                \draw[black, fill=black] (\mcx+\x,\mcy) circle (0.04);
                \draw[line width=.6] (\mcx+\x,\mcy) -- (\mcx+\x,\mcy-0.1);
                \switchy (\mcx+\x,\mcy-0.1)
                \draw[line width=.6] (\mcx+\x,\mcy-0.5) -- (\mcx+\x,\mcy-0.7);
                \draw[line width=.6] (\mcx+\x,\mcy-1.3) -- (\mcx+\x,\mcy-1.5);
                \switchy (\mcx+\x,\mcy-1.5)
                \draw[line width=.6] (\mcx+\x,\mcy-1.9) -- (\mcx+\x,\mcy-2.0);
                \ground (\mcx+\x,\mcy-2.0)
                \transmissiony (\mcx+\x,\mcy-0.7, $\frac{1}{10}\,\lambda$)
            };

        \end{tikzpicture}}
        \label{fig:proposed_matching_network}
    }
    \hfill
    \subfloat[switch unit]{
        \centering
        {
        \small
        \begin{tikzpicture}
            \def\height{0.6};
            
            \draw[color=white] (0,-2.3+0.35) -- (0,0);

            \draw[line width=.6] (-0.2,0) -- (0.4,0);
            \draw[black, fill=black] (0.4,0) circle (0.04);

            \draw[line width=.6] (0.4,\height) -- (0.4,-\height);
    
            \draw[line width=.6] (0.4,\height) -- (0.7,\height);
            \switch (0.7,\height)
            
            \draw[line width=.6] (1.1,\height) -- (2.8,\height);
            
            \draw[line width=.6] (0.4,-\height) -- (0.7,-\height);
            \switch (0.7,-\height)
            
            \draw[line width=.6] (1.1,-\height) -- (1.4,-\height);
    
            \begin{scope}[shift={(0.1,0)}]
                \draw[line width=.6] (1.3,-\height-0.3) -- (1.5,-\height-0.3) -- (2.1,-\height+0.3) -- (2.3,-\height+0.3) -- (2.3,-\height-0.3)
                    -- (2.1,-\height-0.3) -- (1.5, -\height+0.3) -- (1.3,-\height+0.3) -- (1.3,-\height-0.3);

                \draw[line width=.6] (2.3,-\height) -- (2.7,-\height);

                \node[anchor=north, align=center] at (1.8,-\height-0.5) {\color{grey} \footnotesize $180^\circ$ phase shift};
                \draw  [line width=.5pt, anchor=center, dashed, color=grey] (1.1,-\height-0.5) rectangle (2.5,-\height+0.5);
            \end{scope}
            
            \draw[line width=.6] (2.8,\height) -- (2.8,-\height);
            \draw[black, fill=black] (2.8,0) circle (0.04);
            \draw[line width=.6] (2.8,0) -- (3.4,0);

            \draw[line width=.5pt, anchor=center, dashed, color=grey] (-0.0,-1.7) rectangle (3.1,1);

        \end{tikzpicture}}
        \label{fig:proposed_switch_unit}
    }
\caption{Proposed \emph{cost-efficient} hybrid beamformer architecture. Sixteen antenna elements form one tile. Each tile is fed by one RF chain, which we represent by the Thévenin equivalent of the RF chain’s PA. %
The PAs are connected to the antenna elements via (i) a reconfigurable matching network, (ii) a 16-way power splitter, and (iii) reconfigurable switch units, with which one can perform analog beamforming. When implementing multiple tiles in a system, one can additionally perform hybrid digital-analog beamforming.}

\label{fig:circuit_proposed_architecture}
\end{figure*}
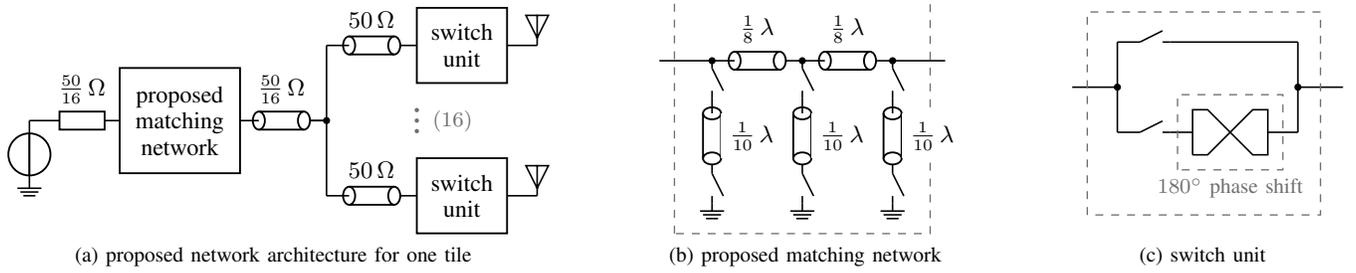

To achieve the objective in~\fref{sec:case_study_objective}, we propose the following hybrid beamformer architecture.
We divide the antenna array into $4\times4$ antenna tiles. Accordingly, the array shown in~\fref{fig:radiating_structure_4} contains one tile, whereas the array in~\fref{fig:radiating_structure_16} contains sixteen tiles. 
Each tile is fed by one PA, whose output impedance is set to $50/16$\,\si{\ohm}.
The tuning network connecting each PA to sixteen antenna elements is shown in~\fref{fig:proposed_tuning_network}. 
It comprises three main components: 
(i) a matching network located directly after the PA, 
(ii) a $16$-way power splitter that distributes the signal to the individual antenna elements, 
and (iii) a switch unit placed ahead of each antenna, replacing traditional, costly phase shifters. 
The proposed matching-network architecture is shown in~\fref{fig:proposed_matching_network}. 
The network is a triple-stub tuner, in which RF switches determine whether each stub is active and whether it is terminated in an open or short circuit. 
Using three stubs provides an acceptable trade-off between high matching efficiency and the additional losses introduced by (realistic and thus, slightly lossy) switches.
The proposed switch-unit architecture is depicted in~\fref{fig:proposed_switch_unit}. 
Each switch unit comprises two RF switches, enabling four states: 
(i) the signal passes through largely unchanged when only the top switch is closed; 
(ii) the signal passes through with a relative phase shift of approximately $180^{\circ}$ when only the bottom switch is closed; 
(iii) the input signal is reflected at an open-circuit when both switches are open; 
and (iv) the signal is reflected at a short-circuit when both switches are closed. 
We use realistic RF-switch models, which introduce ohmic losses, imperfect isolation, and unwanted reflections.\footnote{The switch parameters were obtained from a Cadence Virtuoso simulation of a realistic RF switch.} 
While these effects are significant, they can be modeled exactly with our circuit-theoretic approach.
\subsection{Benchmark Architectures}\label{sec:benchmark_architectures}

\def\ground (#1,#2) {
    \draw[shift={(#1,#2)}, line width=.6] (-0.16,0) -- (0.16,0);
    \draw[shift={(#1,#2)}, line width=.6] (-0.12,-0.05) -- (0.12,-0.05);
    \draw[shift={(#1,#2)}, line width=.6] (-0.08,-0.1) -- (0.08,-0.1);
}

\def\impedanceName (#1,#2,#3) {
    \draw  [shift={(#1,#2)}, line width=.7pt, anchor=center, fill=white] (-.3,-.12) rectangle (.3,.12);
    \node[shift={(#1,#2)}, anchor=south] at (0,0.12) {#3};
}

\def\impedance (#1,#2) {
    \draw  [shift={(#1,#2)}, line width=.7pt, anchor=center, fill=white] (-.3,-.12) rectangle (.3,.12);
}

\def\voltage (#1,#2) {
    \draw[shift={(#1,#2)}, line width=.7pt] (0,0) ellipse (0.28 and 0.28);
    \draw[shift={(#1,#2)}, line width=.7pt] (0,-.28) -- (0,.28);
}

\def\cont (#1,#2,#3) {
    \draw  [shift={(#1,#2)}, line width=.5pt, fill=grey, color=grey] (-.2,.14)  ellipse (0.02 and 0.02);
    \draw  [shift={(#1,#2)}, line width=.5pt, fill=grey, color=grey] (-.2,0) ellipse (0.02 and 0.02);
    \draw  [shift={(#1,#2)}, line width=.5pt, fill=grey, color=grey] (-.2,-0.14) ellipse (0.02 and 0.02);
    \node[shift={(#1,#2)}, anchor=west] at (-.2,0) {\color{grey} \footnotesize #3};
}

\def\antenna (#1,#2) {
    \draw[shift={(#1,#2)}, line width=.6] (0,0) -- (0,0.4) -- (0.15,0.4) -- (0,0.1) -- (-0.15,0.4) -- (0,0.4);
}

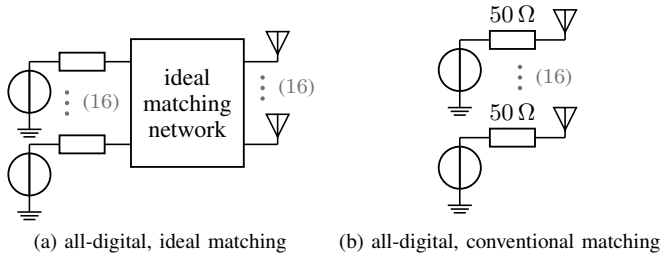
\begin{figure}
    \vspace{-.5cm}
    \centering
    \subfloat[all-digital, ideal matching]{
    \centering
    {
    \small
    \begin{tikzpicture}

        \begin{scope}[shift={(.5,0)}]

        \draw[line width=.6] (2,3.9) -- (5.3,3.9);
        \draw[line width=.6] (2,2.8) -- (5.3,2.8);

        \draw[line width=.6] (2,3.0) -- (2,3.9);
        \draw[line width=.6] (2,1.9) -- (2,2.8);

        \impedance (2.7,3.9)
        \impedance (2.7,2.8)

        \voltage (2,3.5)
        \voltage (2,2.4)

        \ground (2,3.0)
        \ground (2,1.9)

        \cont (2.7,3.35, ($16$))
        
        \cont (5.3,3.55, ($16$))

        \draw  [shift={(4.1,0)}, line width=.7pt, anchor=center, fill=white] (-.75,2.5) rectangle (.75,4.2);
        \node[shift={(4.1,0)}, anchor=center, align=center] at (0,3.35) {ideal\\matching\\network};

        \end{scope}

        \antenna (5.8,3.9)
        \antenna (5.8,2.8)


        \end{tikzpicture}}
        \label{fig:circuit_all_digital_ideal}
    }
    \subfloat[all-digital, conventional matching]{
    \centering
    {
    \small
    \begin{tikzpicture}

        \draw[white, line width=0.0] (.55,2.5-1) -- (4.65,3.9);
        
        \foreach \y in {0, -1.3} {
            \begin{scope}[shift={(0,\y)}]
                \draw[line width=.6] (2,3.9) -- (3.4,3.9);
        
                \draw[line width=.6] (2,3.0) -- (2,3.9);
        
                \impedanceName (2.7,3.9,\SI{50}{\ohm})
        
                \voltage (2,3.5)
        
                \ground (2,3.0)
        
                \antenna (3.4,3.9)
            \end{scope}
        }
        
        \cont (3.0,3.4, ($16$))

        \end{tikzpicture}}
        \label{fig:circuit_all_digital_conventional}
    }
\caption{Architecture of the all-digital with ideal matching and the all-digital with conventional matching benchmark architectures.}

\label{fig:circuit_all_digital}
\end{figure}

In our experiments, we compare the proposed architecture in~\fref{sec:proposed_architecture} to the following benchmark architectures. 
\subsubsection{All-Digital, Ideal Matching}\label{sec:all_digital_with_ideal_matching} 
An all-digital beamformer in which the $N$ PAs are ideally matched to the $M = N$ antenna elements (see~\fref{fig:circuit_all_digital_ideal}) represents the fundamentally best system that can theoretically be realized with a given radiating structure.
Consequently, following~\fref{rem:G_R_explain}, its REMS gain $\hat{G}_\up{REMS}(\theta,\varphi)$ is equal to the radiating gain $\hat{G}_\up{R}(\theta,\varphi)$ of the used radiating structure. 
Following~\fref{rem:gain_inequality}, we use the performance of this all-digital, ideally matched benchmark to quantify how close a given system comes to the fundamentally best achievable performance for the given radiating structure.
\subsubsection{All-Digital, Conventional Matching}\label{sec:all_digital_conventional_matching_architecture}
The ideal benchmark system described above in \fref{sec:all_digital_with_ideal_matching}, with current technologies, is either not practically realizable or only realizable at high costs.
A more realistic benchmark is an all-digital beamformer with conventional matching, i.e., in which the RF chains are directly connected to the antenna array and the PA output impedance is approximately matched to the antenna input impedance, which in our case is close to~\SI{50}{\ohm} (see~\fref{fig:circuit_all_digital_conventional}).

\subsubsection{Proposed Architecture with Ideal Matching}\label{sec:proposed_architecture_ideal_matching}
This benchmark system corresponds to the proposed architecture in~\fref{fig:proposed_tuning_network} with the proposed matching network replaced by an ideal matching network.
Following~\fref{rem:G_T_explain}, its REMS gain $\hat{G}_\up{REMS}(\theta,\varphi)$ is equal to the tuning gain $\hat{G}_\up{T}(\theta,\varphi)$ of the proposed architecture. 
Following~\fref{rem:gain_inequality} and~\fref{rem:G_T_explain}, we use the performance of this benchmark to quantify how close our proposed architecture comes to the fundamentally best achievable performance for the given radiating structure with ideal PA-matching assumed. 

\subsection{Optimization of RF Switch Positions}
Thanks to the efficiency of our REMS model, for the $4 \times 4$ array, the optimal switch positions\footnote{There exist in total $2^{38}$ different combinations for the $4 \times 4$ tile.} can be found in a few days via exhaustive search.
For the $16 \times 16$ array, however, the search space is too large for exhaustive search to remain feasible. 
Thus, we use a heuristic algorithm based on coordinate ascent to determine favorable switch states. 
\section{Results}\label{sec:results}
\begin{figure}[btp]
    \centering
    \includegraphics{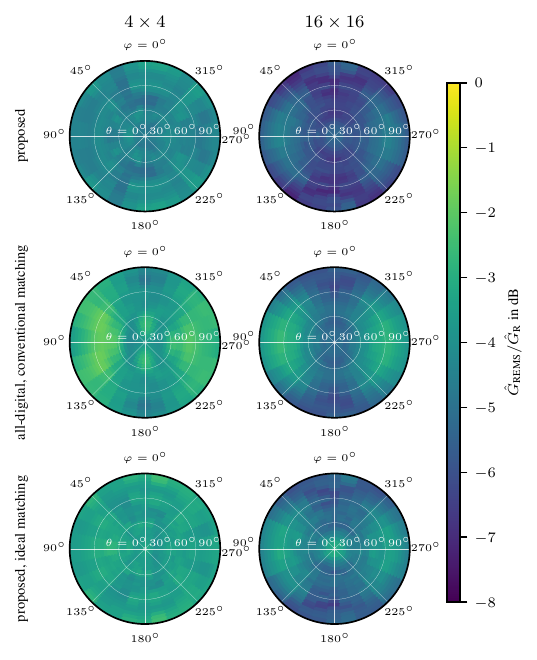}
    
    \caption{Relative REMS gain, $\hat{G}_\up{REMS}(\theta,\varphi) / \hat{G}_\up{R}(\theta,\varphi)$ of the various analyzed architectures.  
    This ratio between REMS gain and radiating gain quantifies how closely each architecture approaches the maximal performance achievable with the given arrays. 
    The results are shown for $(\theta,\varphi) \in [0, \pi/2] \times [0, 2\pi)$.}

\label{fig:plot_gain_rems_t_digital}
\end{figure}

\begin{figure}
    \vspace{-.25cm}
    \centering
    \includegraphics{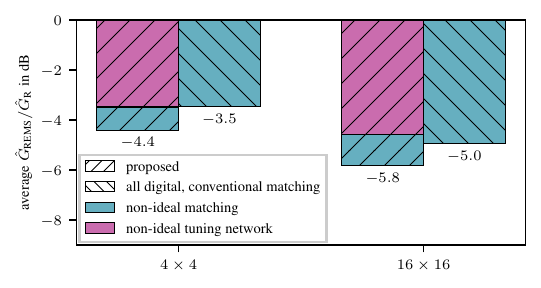}
    \caption{Averaged relative REMS gain, $\hat{G}_\up{REMS}(\theta,\varphi) / \hat{G}_\up{R}(\theta,\varphi)$, for the proposed architecture and for an all-digital, conventionally matched architecture. 
    The average is defined as the median over $(\theta,\varphi) \in [0,\pi/2] \times [0,2\pi)$.
    For the all-digital architecture with a conventional matching network, gain is lost purely due to non-ideal matching. 
    However, for the proposed architecture, we can distinguish between gain losses due to non-ideal matching and those due to the non-ideal tuning network, which includes, among other effects, losses from non-ideal beamforming.
    The proposed architecture achieves only approximately \SI{1}{\decibel} lower gain, while using $16\times$ fewer RF chains.}

\label{fig:plot_average_relative_gain}
\end{figure}

We begin by analyzing the ratio of the REMS gain~$\hat{G}_\up{REMS}(\theta,\varphi)$ to the radiating gain~$\hat{G}_\up{R}(\theta,\varphi)$. As discussed in~\fref{rem:G_R_explain}, this ratio quantifies how closely a system approaches the fundamental performance limit for a given radiating structure and direction $(\theta,\varphi)$. 
We compare three specific systems: (i) the proposed architecture (see~\fref{sec:proposed_architecture}), (ii) the all-digital architecture with conventional matching (see~\fref{sec:all_digital_conventional_matching_architecture}), and (iii) the proposed architecture with ideal matching (see~\fref{sec:proposed_architecture_ideal_matching}).
Furthermore, using the fact that the REMS gain of the proposed architecture with ideal matching corresponds to the tuning gain of the proposed architecture and following~\fref{rem:G_T_explain}, one can quantify the proposed architecture's losses due to non-ideal matching. 
\fref{fig:plot_average_relative_gain} illustrates the average ratio $\hat{G}_\up{REMS}(\theta,\varphi)/\hat{G}_\up{R}(\theta,\varphi)$ for the proposed architecture and the all-digital architecture with conventional matching.
Remarkably, despite using $16\times$ fewer RF chains, the proposed architecture performs within \SI{4.4}{\decibel} (for the $4\times4$ array) and \SI{5.8}{\decibel} (for the $16\times16$ array) of the fundamentally best system (all-digital with ideal matching).
Moreover, when compared to the practical benchmark of an all-digital beamformer with conventional matching, the proposed architecture achieves only approximately \SI{1}{\decibel} lower gain, while again using $16\times$ fewer RF chains.

\begin{figure}
    \vspace{-.25cm}
    \centering
    \includegraphics{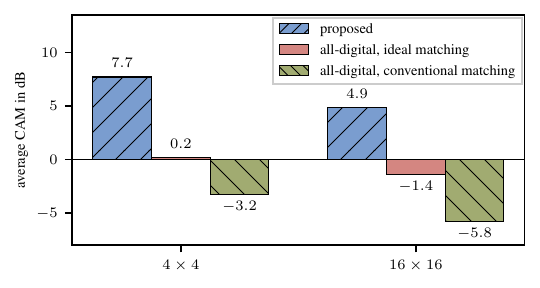}
    \caption{Averaged cost-aware metric $\up{CAM} \triangleq \frac{\hat{G}_\up{REMS}(\theta,\varphi)}{\up{\#~RF~chains}}$ for the proposed architecture, the all-digital, ideally matched architecture, and the all-digital, conventionally matched architecture. 
    The average is defined as the median over~$(\theta,\varphi) \in [0,\pi/2] \times [0,2\pi)$. 
    For both arrays, the proposed architecture offers significantly higher gain per RF chain compared to both of the considered all-digital architectures.}
\label{fig:plot_average_money}
\end{figure}

To further demonstrate the cost efficiency of our architecture, we introduce a cost-aware metric $\up{CAM}(\theta,\varphi) \triangleq \frac{\hat{G}_\up{REMS}(\theta,\varphi)}{\up{\#~RF~chains}}$, which quantifies the achievable gain per RF chain.
In~\fref{fig:plot_average_money}, we visualize the average value of this cost-aware metric for (i) the proposed architecture, (ii) the all-digital architecture with ideal matching, and (iii) the all-digital architecture with conventional matching.
For both arrays, the proposed architecture offers approximately \SI{7}{\decibel} higher gain per RF chain compared to the all-digital architecture with ideal matching and approximately \SI{11}{\decibel} higher gain per RF chain compared to the all-digital architecture with conventional matching.
%

\section{Conclusions}\label{sec:conclusions}

We have applied the \emph{efficient} and \emph{physically consistent} model proposed in~\cite{stutz_schwan_studer_efficient_and_physically_consistent_modeling_of_reconfigurable_electromagnetic_structures} to design and analyze a novel beamforming architecture. 
Specifically, we have proposed a joint hybrid beamforming and matching architecture that relies on cost- and power-efficient RF switches and transmission lines. 
Numerical results based on full-wave EM simulations have demonstrated that the proposed architecture attains a beamforming gain comparable to all-digital architectures with conventional matching, while requiring only a fraction of the expected cost. 
Furthermore, we have shown that our architecture scales favorably with the number of antennas, making it a promising candidate for very large antenna arrays. 

While this case study analyzed an inter-antenna spacing of $\frac{\lambda}{4}$, the trade-off between the added complexity of this antenna spacing relative to a standard $\frac{\lambda}{2}$ array warrants further investigation.
Overall, our results highlight how physically consistent modeling enables the design and analysis of new beamforming architectures, opening up novel opportunities for energy- and cost-efficient large-scale antenna systems.


\balance
\bstctlcite{IEEEexample:BSTcontrol} 
\bibliographystyle{IEEEtran}
\bibliography{bib/publishers,bib/journals_proceedings_ect,bib/library}
\balance

\end{document}